\documentclass[sigconf]{acmart} 
\usepackage{utfsym}
\usepackage{pifont}
\usepackage{multirow}
\usepackage{marvosym}
\usepackage{tcolorbox}
 \usepackage{array}
\usepackage[linesnumbered,ruled,vlined]{algorithm2e}
\usepackage{enumitem}
\usepackage{svg}
\usepackage{subfigure}
\usepackage[doipre={doi:~}]{uri}
\acmConference[ISSTA 2023]{ACM SIGSOFT International Symposium on Software Testing and Analysis}{17-21 July, 2023}{Seattle, USA}
\def\eg{\emph{e.g.}}
\def\ie{\emph{i.e.}}

\newcommand{\etal}{\textit{et} \textit{al}.\space}

\newcommand{\tool}{LIMI\space}
\newcommand{\toolns}{LIMI}

\newcommand{\revised}[1]{\textcolor{black}{#1}}

\AtBeginDocument{%
  \providecommand\BibTeX{{%
    \normalfont B\kern-0.5em{\scshape i\kern-0.25em b}\kern-0.8em\TeX}}}

\setcopyright{acmcopyright}
\copyrightyear{2023}
\acmYear{2023}
\acmDOI{XXXXXXX.XXXXXXX}

\acmPrice{15.00}
\acmISBN{978-1-4503-XXXX-X/18/06}

\begin{document}

\title{Latent Imitator: Generating Natural Individual Discriminatory Instances for Black-Box Fairness Testing}





\author{Yisong Xiao$^{1,2}$, Aishan Liu$^{1,3}$\textsuperscript{\Letter}, Tianlin Li$^{4}$, Xianglong Liu$^{1,3,5}$\textsuperscript{\Letter}} 
\thanks{\Letter\,Corresponding authors.}

\affiliation{\country{$^1$ NLSDE, Beihang University, Beijing, China \quad  $^2$ Shen Yuan Honors College, Beihang University, Beijing, China \\ $^3$ Institute of Dataspace, Hefei, Anhui, China \quad
$^4$ Nanyang Technological University, Singapore  \\ $^5$ Zhongguancun Laboratory, Beijing, China} }

\email{ {xiaoyisong, liuaishan, xlliu}@buaa.edu.cn,
 tianlin001@e.ntu.edu.sg}
 


\renewcommand{\shortauthors}{Yisong Xiao, Aishan Liu, Tianlin Li, and Xianglong Liu}

\begin{abstract}
Machine learning (ML) systems have achieved remarkable performance across a wide area of applications. However, they frequently exhibit unfair behaviors in sensitive application domains (\eg, employment and loan), raising severe fairness concerns. To evaluate and test fairness, engineers often generate individual discriminatory instances to expose unfair behaviors before model deployment. 
However, existing baselines ignore the naturalness of generation and produce instances that deviate from the real data distribution, which may fail to reveal the actual model fairness since these unnatural discriminatory instances are unlikely to appear in practice. 
To address the problem, this paper proposes a framework named \textbf{L}atent \textbf{Imi}tator (\toolns) to generate more natural individual discriminatory instances with the help of a generative adversarial network (GAN), where we imitate the decision boundary of the target model in the semantic latent space of GAN and further samples latent instances on it.
Specifically, we first derive a surrogate linear boundary to coarsely approximate the decision boundary of the target model, which reflects the nature of the original data distribution. Subsequently, to obtain more natural instances, we manipulate random latent vectors to the surrogate boundary with a one-step movement, and further conduct vector calculation to probe two potential discriminatory candidates that may be more closely located in the real decision boundary. Extensive experiments on various datasets demonstrate that our \tool outperforms other baselines largely in effectiveness (\emph{$\times$9.42 instances}), efficiency (\emph{$\times$8.71 speeds}), and naturalness (\emph{+19.65\%}) \revised{on average}. In addition, we empirically demonstrate that retraining on test samples generated by our approach can lead to improvements in both individual fairness (\emph{45.67\% on $IF_r$ and 32.81\% on $IF_o$}) and group fairness (\emph{9.86\% on $SPD$ and 28.38\% on $AOD$}). Our codes can be found on our website \cite{ourweb}.

\end{abstract}


\vspace{-0.1in}
\vspace{-0.1in}


\maketitle

\vspace{-0.15in}
\section{Introduction}
\vspace{-0.05in}
ML systems have been extensively involved in making social-critical decisions, such as employment and loan \cite{chan2018hiring,bono2021algorithmic,berk2021fairness,krittanawong2020machine,mukerjee2002multi}. Despite the promising performance, ML systems can also inadvertently introduce societal biases, which violate the \emph{fairness requirements} of software design. Recent studies have revealed that some ML systems may be discriminatory based on protected attributes (\eg, gender, race). For example, the hiring system used by Amazon \cite{dastin2018amazon} is showing a strong preference for male candidates, while rating the resumes of female candidates with lower scores. Therefore, it is crucial to conduct exhaustive tests to detect fairness violations and further reduce discrimination, which can relieve the fairness concerns of ML systems in social-critical applications.

To evaluate fairness, extensive studies have been proposed to generate test samples that can violate the fairness requirement and reveal the discrimination of ML systems. Generally, existing studies can be primarily divided into two categories: individual fairness  \cite{dwork2012fairness,galhotra2017fairness} \revised{evaluation} and group fairness \cite{feldman2015certifying,hardt2016equality} \revised{evaluation}. Individual fairness requires similar decisions for similar individuals, while group fairness requires equal treatment for different groups (grouped by a protected attribute). 
\revised{Individual fairness is capable of identifying discriminatory behaviors that may be ignored by group fairness measures \cite{galhotra2017fairness}. Specifically, individual fairness allows for a more powerful and granular examination of discriminatory behavior that arises due to changes in only the protected attribute in various contexts, while group fairness measures may fail to detect discrimination when the model treats the same group oppositely in different situations \cite{zhang2021efficient}.}
\revised{Therefore, our primary objective in this work is to conduct individual fairness testing, which entails \emph{generating individual discriminatory instances} \footnote{We use ``instance'' or ``sample'' interchangeably.} for testing purposes.}
In other words, we aim to generate two similar instances that only differ in the protected attribute but show different model decisions, which make a system exhibit individual discrimination \cite{galhotra2017fairness,aggarwal2019black}.

For individual fairness testing, existing methods primarily generate discriminatory instances in a two-phase generation framework \cite{galhotra2017fairness,udeshi2018automated,aggarwal2019black,zhang2020white,zhang2021efficient,zheng2022neuronfair,fan2022explanation}. Specifically, they first construct a set of individual discriminatory instances as global seeds via random sampling or gradient-based search; they then conduct an iterative process to search for instances near the global seeds; finally, they utilize these identified discriminatory instances to improve the individual fairness of the tested model via retraining. However, the naturalness of generated instances remains unsatisfactory (\ie, the generated instances often deviate from the original data distribution), which impedes fairness testing in practical application. 
\revised{For example, the generated instances by these methods may violate real-world constraints (\eg, authorize a loan to a 10-year-old individual) \cite{chen2022fairness}, thus falsely estimate the fairness of models.}

To address the problem, this paper proposes a framework named Latent Imitator (\toolns), which can generate natural discriminatory instances that is closer to the original data distribution. Our \tool approach imitates the decision boundary of the target model and probes latent samples on it to generate more natural discriminatory instances. 
\revised{Previous studies \cite{mickisch2020understanding,karimi2019characterizing} have shown that the decision boundary tends to be close to the majority of natural data points after training, which indicates a close alignment between the decision boundary and the distribution of the training data. Therefore, test cases near the decision boundary are more likely to possess better naturalness and have a higher potential to induce discrimination.}
However, for the real-world black-box models, we have no access to the detailed model information and fail to derive a semantic decision boundary in the original input domain. Thus, we first approximate a \textbf{surrogate decision boundary} in the latent space of a generative model, which enjoys semantic property \cite{radford2015unsupervised}. Specifically, we construct an auxiliary dataset that maps the latent space to the model decision space and then learns a linear hyperplane based on the dataset as a surrogate boundary of the real decision boundary. The surrogate decision boundary depicts a coarse region for test sample generation, and instances near it will \revised{possess} better discrimination and naturalness. However, there may exist a gap between the surrogate boundary and the real decision boundary, which may induce an inaccurate latent location. We thus design the \textbf{latent candidates probing} strategy, in which we first locate a latent vector at the surrogate boundary with a one-step movement and then probe two potential candidates on either side of it that may be closer to the real boundary.
Therefore, we can finely probe discriminatory latent vectors on the real decision boundary for better naturalness.

To evaluate the performance of our \toolns, we conduct extensive experiments under multiple datasets on both traditional ML models and novel DNNs. Compared to 6 state-of-the-art individual fairness testing approaches, our \tool generates \emph{$\times$9.42} more discriminatory instances at \emph{$\times$8.71} faster speed, and the distribution of generated instances \revised{possesses} better naturalness (\emph{+19.65\%}) \revised{on average}.
In addition, benefiting from the naturalness of our generated discriminatory instances, we can improve the model fairness in both individual fairness (\emph{45.67\% on $IF_r$ and 32.81\% on $IF_o$}) and group fairness (\emph{9.86\% on $SPD$ and 28.38\% on $AOD$}) via retraining.
Moreover, to better understand our framework, we conduct further investigations and demonstrate that our framework performs well on image datasets and can be flexibly combined with other baselines as a fast global prober.
Our main contributions are:
\begin{itemize} [leftmargin=*]
\item To the best of our knowledge, we are the first to generate natural individual discriminatory instances for fairness testing, which could help reveal the actual model fairness.  
\item We propose \tool framework that can generate more natural individual discriminatory instances by imitating the decision boundary of the target model and further sampling latent instances on it.
\item Extensive experiments on several benchmarks demonstrate \tool outperforms SOTA baselines largely in effectiveness (\emph{$\times$9.42 instances}), efficiency (\emph{$\times$8.71 speeds}), and naturalness (\emph{+19.65\%}) \revised{on average}.
\item We publish \tool as a self-contained toolkit on our website \cite{ourweb}.
\end{itemize}

\vspace{-0.15in}
\section{Preliminaries}
\vspace{-0.05in}
In this section, we first provide a brief review of the relevant background, including the binary classification model, and generative adversarial networks; we then illustrate the problem definition.
\vspace{-0.1in}
\subsection{Backgrounds and Notations}

\quad \textbf{\emph{Binary Classification Model}}. Given a dataset $\mathcal{D}$ with data sample $x \in X$ and label $y \in Y$, the binary classification model can be represented as $f(x,\theta):X \rightarrow Y$, where $Y$ only contains two classes and $\theta$ denotes the parameters of model $f$. The goal of these binary classification models (\eg, Random Forests (RF) \cite{ho1998random}, Support Vector Machines (SVM) \cite{cortes1995support}, and Deep neural networks (DNN) \cite{lecun2015deep}) is to learn a decision boundary, which is a surface in the input domain that can separate input instances into two classes. Formally, the decision boundary can be written as $F = \{x: \theta(x)=0\}$, where $\theta(x)<0$ denotes $x$ is classified as the negative class, and $\theta(x)>0$ is the positive one.

\textbf{\emph{Generative Adversarial Networks (GANs)}}. GANs are commonly used to generate high-quality samples for data synthesis. The GAN framework was first proposed for synthetic images by Goodfellow \etal  \cite{NIPS2014_5ca3e9b1}, which plays a zero-sum game between the generator $G$ and the discriminator $D$. Generator $G$ is trained to mimic the given real training dataset distribution, which maps a synthetic sample $g(\mathbf{z})$ from random latent vector $\mathbf{z}$; while discriminator $D$ is trained to distinguish the generated samples from real samples. The core idea of GAN is the adversarial training process according to the following min-max objective:
\begin{equation}
\mathop{\min}\limits_{G}\mathop{\max}\limits_{D} \mathbb{E}_{x \sim p_{data}}[\log D(x)] + \mathbb{E}_{\mathbf{z} \sim p_{\mathbf{z}}}[\log (1-D(G(x)))],
\end{equation}
where $x \sim p_{data}$ denotes the real distribution, and $\mathbf{z} \sim p_{\mathbf{z}}$ denotes the distribution of latent space. 

Besides visual images, GANs can also be utilized for tabular data generation. Xu \etal \cite{xu2018synthesizing} first proposed TGAN to map the continuous noise latent into discrete tabular data. Based on that, Xu \etal \cite{xu2019modeling} proposed CTGAN, which invents the mode-specific normalization to overcome the non-Gaussian and multi-modal distribution, and further designs to deal with the imbalanced discrete columns.

\vspace{-0.1in}
\subsection{Problem Definition}
\quad \textbf{\emph{Individual discrimination.}} 
Let $A = \{A_1,A_2,...,A_n\}$ and $I=\{I_1,I_2,...,I_n\}$ represent the attributes set of data sample\revised{s} $X$ and its input domain. Following \cite{zhang2020white}, individual discrimination refers to the unequal treatment (\eg, different model decisions) of two similar individuals who only differ in the protected attributes. 
\revised{Here, the protected attributes (\eg, gender, race, and age) have been clearly defined and recognized in various laws and policies (\eg, GDPR and UK equality law). 
Following the commonly-adopted settings  \cite{galhotra2017fairness,udeshi2018automated,aggarwal2019black,zhang2020white,zhang2021efficient,zheng2022neuronfair,fan2022explanation,udeshi2018automated}, the protected attributes are assumed to be a known prior, and we use $PA$ to denote the set of protected attributes and $NPA$ to denote the non-protected attributes.}
For each instance $x = \{x_1,x_2,...,x_n\} \in X$ ($x_i$ is the value of corresponding attribute $A_i$), we define that $x$ is an individual discriminatory instance for model $f$ when there exists another instance $x'$ that satisfies
\begin{equation}
\begin{aligned}
    f(x,\theta) &\neq f(x^{'},\theta) \\
s.t. \exists p \in \revised{PA}, x_p \neq &x_p^{'};  \forall  q \in \revised{NPA}, x_q = x_q^{'}
\end{aligned}
\end{equation}

Moreover, we use $(x,x^{'})$ to represent an individual discriminatory instance pair for $f$. Thus, in this paper, we aim to generate \textbf{individual discriminatory instance} for fairness testing, which could force the target model to produce biased decisions and violates the individual fairness requirements. 
\revised{
In accordance with common assumptions in black-box fairness testing \cite{udeshi2018automated,aggarwal2019black,fan2022explanation}, we assume that the data samples $X$, the corresponding feature space $A$, and the protected attributes set $PA$ are provided for testers, while the target model $f$ trained on dataset $\mathcal{D}=\{X,Y\}$ is black-box (without prior knowledge of inner information like gradients).}

\revised{
\textbf{\emph{Naturalness of generated instances.}} 
}
\revised{In addition to individual discrimination, we also prioritize the naturalness of generated discriminatory instances. 
In the context of tabular data, naturalness is a measure of how well the dataset reflects the underlying distribution of the population \cite{3DQ}. Specifically, naturalness refers to how closely the statistical patterns and relationships in a dataset, such as value distributions and attribute correlations, match the distribution of the population it represents \cite{brenninkmeijer2019generation}.
To quantify the naturalness of generated data, statistical measures such as the Average Nearest Neighbor Distance \cite{peterson2009k}, Pearson’s correlation \cite{sedgwick2012pearson}, and Kolmogorov-Smirnov statistic \cite{massey1951kolmogorov} can be used. These metrics evaluate the distributional distance/similarity between the original and generated data, with the original data (real-world datasets) serving as the reference population. Generated data with high naturalness has a closer distribution to the original dataset, indicating that the data is a more accurate representation of the population \cite{wen2021causal,xu2019modeling}.
Therefore, we aim to generate individual discriminatory instances with high naturalness, which are more likely to appear in the population and can better reveal unfair behaviors in the real world.}
\begin{figure*}[ht]
\centering
\includegraphics[width=0.95\linewidth]{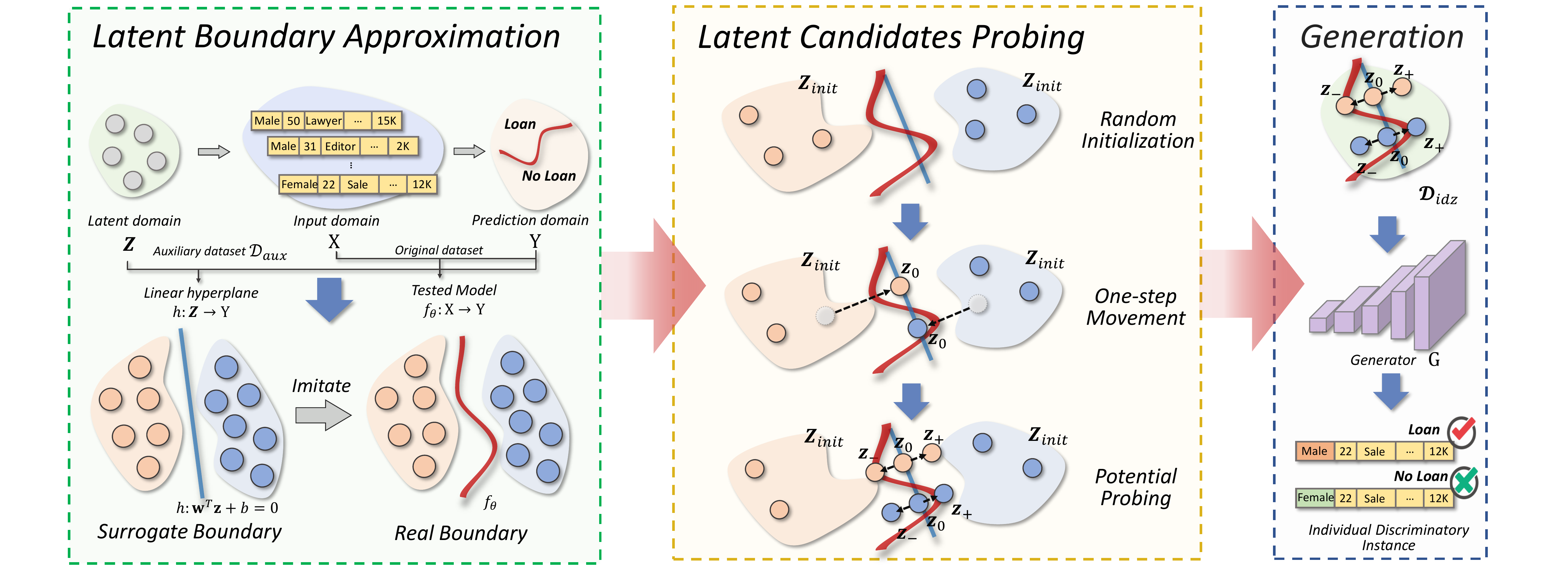}
\vspace{-0.15in}
\caption{Overview of \tool framework. 
\tool first imitates the target model's decision boundary by deriving a surrogate boundary based on $\mathcal{D}_{aux}$, which coarsely depicts a region with better naturalness. Later, to obtain more natural samples, \tool manipulates random latent vectors to the surrogate boundary with a one-step movement and further probes two potential discriminatory candidates that may be closer to the real boundary.
}

\label{fig:framework}
\vspace{-0.2in}
\end{figure*}

\vspace{-0.1in}
\section{Methodology}

In this paper, we propose a framework named \textbf{L}atent \textbf{Imi}tator (\toolns) to generate natural individual discriminatory instances for black-box fairness testing. Since the decision boundary of the target model often reflects the nature of the original data distribution, we first coarsely approximate the decision boundary of the target model by deriving a surrogate boundary in the latent space of GAN. Thus, test cases near the surrogate boundary are more likely to serve as discriminatory instances with slight perturbations and \revised{possess} better naturalness. However, there may exist a gap between the surrogate boundary and the real boundary, which would hinder the accurate location of latent sample generation. 
Therefore, we then manipulate random latent vectors with a one-step movement to the surrogate boundary and probe two potential candidates, so that we could find potential discriminatory latent vectors that may be more closely located around the decision boundary leading to better naturalness. The overall framework is shown in Fig \ref{fig:framework}.

\vspace{-0.1in}
\subsection{Latent Boundary Approximation}
\vspace{-0.01in}

\revised{As established in previous work \cite{mickisch2020understanding,karimi2019characterizing}, the decision boundary of the model is often close to the majority of natural data points after training, which highly reflects the nature of the original data distribution at the same time.} 
Therefore, test cases generated under the guidance of the target model decision boundary could be obviously more natural \revised{(\ie, closer to the original distribution)}.
However, in the black-box fairness testing scenario, we cannot directly access the model and fail to derive a semantic decision boundary in the original input domain. 
Thus, we try to approximate the target model's decision boundary by deriving a surrogate boundary in the latent space of GAN based on a constructed synthetic latent dataset.

GAN has been widely used to synthesize massive realistic data for software testing \cite{zhang2018deeproad,gao2019automated}, therefore we use GANs to help us generate the synthetic latent dataset. Specifically, we can derive a sample $x$ in the input domain from a randomly sampled latent vector $\mathbf{z}$ by the generator $G$ of GAN as $g(\mathbf{z}): \mathbf{Z} \rightarrow X$; the sample $g(\mathbf{z})$ can be then classified by the target model with a binary label $f(g(\mathbf{z}))$ in the prediction domain. The pipeline implicitly embodies the model decision process on the latent space, which can be used for the latent decision boundary approximation. Based on the above analysis, we aim to find a surrogate decision boundary $F_{latent}=\{\mathbf{z}:\theta(g(\mathbf{z}))=0\}$ in the latent space to characterize the implicit relationship, so that we can approximate the real decision boundary of the target black-box model for subsequent individual discriminatory instances generation.
Since studies \cite{denton2019image,shen2020interpreting} have illustrated that the latent space of GAN is approximately linearly separable in any binary semantic attribute (\eg, whether to loan), we, therefore, employ a linear hyperplane as the coarse surrogate decision boundary. Coarsely, $F_{latent}$ can be written as following expression
\begin{equation} 
h: \mathbf{w}^{T} \mathbf{z} + b=0,
\end{equation} 
where $w$ denotes the normal vector, and $b$ represents the intercepts. 

The latent vector $\mathbf{z}$ is assigned to an \revised{unfavorable} sample (\eg, not loan) of model $f$ when it satisfies $\mathbf{w}^T \mathbf{z} + b<0$, otherwise, a \revised{favorable} sample (\eg, loan) if it satisfies $\mathbf{w}^T \mathbf{z} + b>0$. It is worth noting that unlike a surrogate decision tree produced by LIME \cite{ribeiro2016should} in the original input domain, the surrogate boundary in the continuous latent space has vector arithmetic property  \cite{radford2015unsupervised}, which indicates that latent vectors can be manipulated semantically under its guidance. We can easily obtain a sample of the opposite class from the initial random point through latent vector arithmetic (\eg, adding with the normal vector $w$). 
Therefore, the operable (\ie, vector arithmetic property) surrogate boundary establishes a foundation for the subsequent probing process.

To better obtain the surrogate decision boundary (\ie, a linear hyperplane), we need to construct an auxiliary latent dataset, which maps the latent space to the model decision space. Concretely, we sample a set of latent vectors $\mathbf{Z}_{init}$ randomly and then utilize the generator of GAN to synthesize samples in the original input domain $X$; we then label each sample with the corresponding prediction result based on the target tested model. Formally, the auxiliary latent dataset $\mathcal{D}_{aux}$ can be expressed as
\begin{equation} 
\mathcal{D}_{aux}=\{(\mathbf{z},f(g(\mathbf{z}))):\mathbf{z} \in \mathbf{Z}_{init} \}. 
\end{equation} 

Note that the fitness of our surrogate boundary is highly related to the effectiveness of subsequent test case generation. Thus, we try to obtain a more precise boundary by refining the auxiliary dataset $\mathcal{D}_{aux}$ with the ranking and resampling process. 
\revised{
    Since high-confidence samples tend to be more concentrated, while low-confidence samples are often distributed at the twists and turns of the decision boundary, denser high-confidence samples have a higher probability of reflecting the real condition of the lumpy neural network boundary \cite{guan2020analysis}. 
}
Therefore, we first rank the auxiliary latent dataset based on each classification score assigned by model $f$, then retain latent vectors whose scores are higher than threshold $\epsilon$.
To guarantee sufficient high-quality training samples, $\epsilon$ is empirically set as 0.7.
Considering the imbalance of the synthetic data, we adopt a random over-sampling strategy \cite{santoso2017synthetic} that adds more samples from the minority class in $\mathcal{D}_{aux}$ by simply replicating, so as to construct a class-balanced dataset.
Based on the constructed auxiliary dataset $\mathcal{D}_{aux}$, we finally train our linear representation model (\ie, SVM with linear kernel) to derive a surrogate decision boundary $F_{latent}$. 

To sum up, we first approximate the boundary of the target model by deriving a surrogate decision boundary in the latent space of GAN based on a constructed synthetic latent dataset. The surrogate decision boundary depicts the target region for the generation of individual discriminatory instances (\ie, \revised{possess} both natural and discriminatory properties) in the latent space coarsely, which will provide strong guidance for the later latent probing process.

\vspace{-0.1in}
\subsection{Latent Candidates Probing} \label{subsection: LCP}
\vspace{-0.02in}

Existing studies  \cite{zhang2020white,moosavi2016deepfool,brendel2017decision} have revealed the fact that data points near the decision boundary are more likely to mislead the model's predictions where only a slight perturbation is required. Therefore, after obtaining a latent surrogate decision boundary, we probe the latent space for the random initial samples and move them near the surrogate boundary, so that we could find potential natural and discriminatory latent vectors. 

With the surrogate boundary $h$ obtained in latent space, we can easily calculate the distance $d$ from any random latent vector $\mathbf{z}$ to $h$, which reflects the decision score of model $f$ under the corresponding test case $x$:
\begin{equation} 
d =  \frac{\mathbf{w}^{T}\mathbf{z}+b}{\sqrt{\mathbf{w}^T\mathbf{w}}}. \label{distance_equation}
\end{equation}

When a latent vector $\mathbf{z}$ gets closer to the surrogate boundary, the uncertainty of model prediction increases extremely, while the corresponding synthetic sample still stays in the central distribution of the original dataset. Therefore, when we generate a test case $x$ from latent $\mathbf{z}$ near our surrogate boundary and modify its protected attribute value to get another instance $x^{'}$, they will most likely get different prediction results from the target model. This assumption exploits the inherent adversarial robustness property of machine learning models \cite{szegedy2013intriguing}, that is a model can be misled by a slight perturbation added to the input. In this way, we can obtain an individual discriminatory instance with maximum possibility and \revised{acquire} naturalness simultaneously.

Based on the above analysis, given a random latent vector $\mathbf{z}$ as an initial point, we can naively make a search process that perturbs it toward the surrogate boundary iteratively, which is similar to the existing fairness testing methods \cite{zhang2020white,zhang2021efficient,zheng2022neuronfair}. The perturbing process can be simply defined as
\begin{equation} 
\mathbf{z}_{next} = \mathbf{z} + dir \cdot s_{p} \cdot \mathbf{w}_{u}, 
\end{equation} 
where $dir$ is the direction ($\pm 1$) to the surrogate boundary, $s_p$ denotes the step size of the iterative search process, and $\mathbf{w}_{u}$ represents the unit of $\mathbf{w}$ (\ie, $\mathbf{w}_{u}=\frac{\mathbf{w}}{\sqrt{\mathbf{w}^T\mathbf{w}}}$), which directs the shortest path.

However, we note the tedious nature of the iterative search process that it is time-consuming which would highly influence the speed of test case generation. Thus, we aim to accelerate the generation process, where we replace the iterative searching process with the one-step probing strategy. The refined latent manipulation can be formally written as
\begin{equation} 
\mathbf{z}_{0} = \mathbf{z} -(\mathbf{w}_{u}^{T} \mathbf{z} + b_{u}) \cdot \mathbf{w}_{u}, 
\end{equation} 
where $b_{u} = \frac{b}{\sqrt{\mathbf{w}^T\mathbf{w}}}$ is also the normalized unit. Due to the implicit constraints of the surrogate boundary in latent space, the manipulated latent vectors still maintain their naturalness. We can verify the property of $\mathbf{z}_0$ ($d(\mathbf{z}_{0})=0$) by taking it into the distance function. Though the single-step probing strategy facilitates efficiency significantly, it may lose accuracy due to the coarseness of our surrogate boundary.

Notice that, in the first coarse step of our approach, we utilize a simple linear hyperplane to approximate tested ML models, while the real decision boundary of these models will be nonlinear and more complex. For example, the boundary of a simple neural network is a subset of tropical hypersurface \cite{alfarra2022decision} due to its non-linear activation. Therefore, the manipulated latent vector $\mathbf{z}_0$ may not accurately locate at the real decision boundary, which fails to reflect strong naturalness. To solve this potential problem, we propose an additional latent candidate probing strategy where we introduce two candidates on the side of $\mathbf{z}_0$ to bridge the approximation gap between our surrogate boundary and the real decision boundary, therefore, leading to better naturalness of the generated instances. Specifically, these two candidates can be calculated with the following expressions
\begin{equation} 
\begin{aligned}
&\mathbf{z}_{+} = \mathbf{z}_{0} + \lambda \cdot \mathbf{w}_{u}, \\
&\mathbf{z}_{-} = \mathbf{z}_{0} - \lambda \cdot \mathbf{w}_{u}, 
\end{aligned}
\end{equation} 
where $\lambda$ represents a hyperparameter that controls the walking length along direction $\mathbf{w}_{u}$. For concrete tabular data, small $\lambda$ would result in generating duplicate test cases, making the candidates useless; on the contrary, big $\lambda$ can incur those candidates to move far away from decision boundaries, which deviates from the original data distribution. We will conduct ablation studies in Section \ref{sec:discussion}.

Therefore, given a set of random latent vectors $\mathbf{Z}_{init}$, for each $\mathbf{z} \in \mathbf{Z}_{init}$ we can calculate potential individual discriminatory latent vectors near the model decision boundary to construct $\mathcal{D}_{idz} =  \{(\mathbf{z}_0,\mathbf{z}_+,\mathbf{z}_-): \mathbf{z} \in \mathbf{Z}_{init} \}$. 

 To sum up, we design two probing strategies to obtain latent vectors that are most likely to generate natural discriminatory test cases, one is to probe latent vectors located at the surrogate boundary, and another is to probe two candidates to bridge the boundary approximation gap. With only a few lightweight vector calculations, we can finely probe many latent vectors that are distributed near the surrogate boundary, which will benefit the naturalness of generated discriminatory instances.

\vspace{-0.1in}
\subsection{Overall Testing Process}

\IncMargin{1em}
\begin{algorithm}[ht]
\footnotesize
    \caption{Latent Imitator}
    \label{algo_all} 
    
\SetKwInOut{Input}{Input}
\SetKwInOut{Output}{Output}
\SetKwFunction{Rank}{Rank}\SetKwFunction{Resample}{Resample}

\Input{model $f_\theta$, generator $G$, $\lambda$, $\epsilon=0.7$} 
\Output{discriminatory instance set $\mathcal{D}_{idi}$}
\BlankLine 

 $\mathbf{Z}_{init}, G(\mathbf{Z}_{init}) \leftarrow$ randomly sample 1M samples from $G$\;
 \tcp{Latent Boundary Approximation}
 $\hat{Y},\hat{S} \leftarrow $    $f_\theta(G(\mathbf{Z}_{init}))$; \tcp{model predictions and scores} 
 
 $index_{0}, index_{1} \leftarrow $ \Rank{$\hat{Y},\hat{S}$}; \tcp{ sort $\hat{S}$ from high to low and omit $\hat{s} < \epsilon$; and $index_{0}$ records $\hat{y}=0$, $index_{1}$ records $\hat{y}=1$}

$index_{0} \leftarrow$ \Resample{$index_{0},50K$}; \tcp{randomly over-sample in $index_{0}$ until 50K}
$index_{1} \leftarrow$ \Resample{$index_{1},50K$}; \tcp{randomly over-sample in $index_{1}$ until 50K}

$\mathcal{D}_{aux} \leftarrow $ $(\mathbf{Z}_{init}[index_{0}],0) \cup (\mathbf{Z}_{init}[index_{1}],1)$; \tcp{construct the auxiliary dataset}
 
 train the surrogate decision boundary $h$ based on $\mathcal{D}_{aux}$; \tcp{implemented by a linear SVM, $h: \mathbf{w}^{T} \mathbf{z} + b=0$}

 \tcp{Latent Candidates Probing}
\revised{$\mathcal{D}_{idz} \leftarrow \emptyset$\;}

\ForEach{$\mathbf{z}$ \rm{in} $\mathbf{Z}_{init}$}{
    
    $\mathbf{z}_{0} \leftarrow \mathbf{z} -(\mathbf{w}_{u}^{T} \mathbf{z} + b_{u}) \cdot \mathbf{w}_{u}$\;
    $\mathbf{z}_{+} \leftarrow \mathbf{z}_{0} + \lambda \cdot \mathbf{w}_{u}$\;
    $\mathbf{z}_{-} \leftarrow \mathbf{z}_{0} - \lambda \cdot \mathbf{w}_{u}$\; 
    \revised{$\mathcal{D}_{idz} \leftarrow \revised{\mathcal{D}_{idz}} \cup (\mathbf{z}_{0}, \mathbf{z}_{+}, \mathbf{z}_{-})$}
    
}
\revised{\tcp{Generation}}
\revised{
$\mathcal{D}_{test} \leftarrow G(\mathbf{Z}_{idz})$; \tcp{constructed as Equation (\ref{equation:Dtest})}
}
\revised{$\mathcal{D}_{idi} \leftarrow \emptyset$\;}
\revised{
\ForEach{($g(\mathbf{z}_{0}),g(\mathbf{z}_{+}),g(\mathbf{z}_{-}))$ \rm{in} $\mathcal{D}_{test}$}{
    $x_{0},x_{+},x_{-} \leftarrow g(\mathbf{z}_{0} ),g(\mathbf{z}_{+} ),g(\mathbf{z}_{-} )$\;
    \ForEach{$x$ \rm{in} $(x_{0},x_{+},x_{-})$ }{
        \If{ $x$ is a individual discriminatory instance}{
            $\mathcal{D}_{idi} \leftarrow \mathcal{D}_{idi} \cup x$\;
            break\;
        }
    }
}
}
return $\mathcal{D}_{idi}$\;

\end{algorithm}

\DecMargin{1em}

Algorithm \ref{algo_all} shows the overall testing process of our proposed \tool approach. In general, given a black-box target model $f_\theta$ and a generator $G$ both trained on dataset $X$, our testing framework can be roughly divided into two phases, \ie, latent boundary approximation and latent candidates probing. In the first phase, we construct an auxiliary dataset $\mathcal{D}_{aux}$ that consists of latent vectors and corresponding decision labels from model $f_\theta$, and then adopt a linear kernel SVM to derive a surrogate decision boundary in the latent space based on $\mathcal{D}_{aux}$. In the second phase, we manipulate random latent vectors to the surrogate boundary in one step and probe two candidates near them for better locating the real decision boundary, and then collect the probed latent vectors to construct potential discriminatory vectors $\mathcal{D}_{idz}$.

After obtaining the latent candidates, we utilize the generator $G$ to synthesize $\mathcal{D}_{idz}$ as the set of potential individual discriminatory instances as
\begin{equation} 
\label{equation:Dtest}
\mathcal{D}_{test}=\{(g(\mathbf{z}_0),g(\mathbf{z}_{+}),g(\mathbf{z}_{-})):(\mathbf{z}_0,\mathbf{z}_+,\mathbf{z}_-) \in \mathcal{D}_{idz}\}.
\end{equation} 

Finally, we check whether a test case in $\mathcal{D}_{test}$ is a discriminatory instance by modifying the value of the selected protected attribute. Once a candidate sample in the tuple $(g(z_0),g(z_{+}),g(z_{-}))$ is determined as a discriminatory instance, we will add it into the discriminatory set $\mathcal{D}_{idi}$ and break out to stop the testing of the left samples in this tuple immediately.

It is worth noting that our approach has no iterative search process, which is much more efficient than the search-based generation framework of other approaches \cite{zhang2020white,zhang2021efficient,zheng2022neuronfair}. Literally, the search-based framework needs to compute guidance (\ie, direction to decision boundary) before each iteration, while our method avoids it due to the latent vector arithmetic property \cite{radford2015unsupervised}. Though researchers have been optimizing the guidance computation methods (\eg, modifying the loss function of gradient calculations) to improve search efficiency in their long-term work, the search process still consumes a long time. In our framework, the surrogate decision boundary depicts a probing region coarsely, then the tedious iteration search is replaced with a latent probing process, which significantly accelerates the testing process. 

\vspace{-0.1in}
\section{Evaluation}
\label{sec:evaluation}
In this section, we evaluate the performance of \tool in terms of effectiveness, efficiency, the naturalness of generated instances, and the utility for fairness improvement. We first outline the experimental setup and then conduct the evaluation by answering the following research questions.

\textbf{RQ1}: How effective and efficient is \tool in finding individual discriminatory instances?

\textbf{RQ2}: How natural are the individual discriminatory instances generated by \toolns?

\textbf{RQ3}: How useful are the generated individual discriminatory instances for improving the model's individual and group fairness?

\vspace{-0.1in}
\subsection{Experimental setup}
\subsubsection{Datasets} We evaluate \tool on four public tabular datasets, which are widely adopted in the fairness testing literature \cite{zhang2020white,udeshi2018automated,aggarwal2019black}. The following is a brief description of these datasets. 
\begin{itemize}[leftmargin=*]
\item \emph{Adult Income Dataset} \cite{adult2017} is used for income prediction with \revised{48,842 samples including 32,561 train samples (7,841 favorable vs 24,720 unfavorable)} and \revised{16,281 test samples (3,846 favorable vs 12,435 unfavorable)}, and the label denotes whether an individual’s annual income is over $\$50K$. The protected attributes in 13 attributes \revised{(5 numerical and 8 categorical)} of Adult are gender, race, and age respectively.

\item \emph{German Credit Dataset} \cite{credit2017} is used for credit risk level prediction (i.e., good or bad) with 600 samples \revised{(300 favorable vs 300 unfavorable)} and 20 features \revised{(7 numerical and 13 categorical)}. The protected attributes are gender and age.

\item \emph{Bank Marketing Dataset} \cite{bank2017} labels whether a client will subscribe to a term deposit or not. There are \revised{45,211 samples (5,289 favorable vs 39,922 unfavorable)} and 16 features \revised{(6 numerical and 10 categorical)}, and the protected attribute is age.

\item \emph{Medical Expenditure Panel Survey Dataset} \cite{meps2015} is used for health care needs prediction with \revised{15,675 samples (2,628 favorable vs 13,047 unfavorable)} and 40 features \revised{(4 numerical and 36 categorical)}, and the protected attribute is gender.
\end{itemize}
We use Adult, Credit, Bank, and Meps for simplicity. 
\revised{
To ensure the rationality of subsequent analysis, we also examine the correlation of attributes using Spearman’s rank order method \cite{spearman1961general}. We find that the protected attributes are not significantly correlated to any other attributes (average correlation among attributes, Adult: 0.09, Credit:0.10, Bank:0.06, Meps:0.05), which is consistent with the observations in \cite{zhang2021efficient}.
}
\vspace{-0.05in}
\subsubsection{Binary Classification Models} We employ three commonly-used binary classification models to evaluate \toolns. The traditional machine learning models are RF and SVM implemented in scikit-learn \cite{kramer2016scikit} library. 
Especially, the RF consists of 100 trees, and the SVM is implemented with an RBF kernel, following \cite{fan2022explanation}.
The DNN model is a six-layer fully-connected neural network, with architecture \cite{zhang2020white} in detail. And we optimize the DNN models for 1000 epochs by the Adam optimizer with a learning rate of 0.001. These models all perform well in classification, with an average accuracy of 91.60\%. 
\revised{Our training and testing settings align with commonly-adopted practices in \cite{fan2022explanation,zhang2020white,zhang2021efficient,zheng2022neuronfair}, and we provide more information on our implementation and results on our website \cite{ourweb}.}

\vspace{-0.05in}
\subsubsection{Baselines} 
\revised{We compare \tool with} 6 state-of-the-art approaches, which can be divided into two types based on the tested model. \revised{The first type} includes Aequitas \cite{udeshi2018automated}, SymbGen \cite{aggarwal2019black}, and ExpGA \cite{fan2022explanation}, which provide the testing ability for all machine learning models. \revised{The second type} includes ADF \cite{zhang2020white}, EIDIG \cite{zhang2021efficient}, and NeuronFair \cite{zheng2022neuronfair}, which are designed especially for DNN.
\revised{We use the implementation of these approaches directly from corresponding GitHub repositories and adhere to the best settings reported in their papers.}

\vspace{-0.05in}
\subsubsection{Evaluation Metrics} \label{subsubsec:Evaluation-Metrics}
The following three aspects of \tool are evaluated, including the effectiveness and efficiency of generation, the naturalness of generated individual discriminatory instances, and the fairness improved after retraining.

\textbf{Effectiveness and efficiency Evaluation}. We use $\#\mathcal{D}_{idi}$ and $EGS$ metrics to measure the effectiveness and efficiency of the individual discriminatory instances generation procedure. $\# \mathcal{D}_{idi}$ denotes the absolute quantity of generated individual discriminatory instances. $EGS$ denotes the speed of effective test case generation (\ie, how many individual discriminatory instances can the approach generate per second):
\textcolor{black}{\small{
\begin{equation} 
EGS = \frac{\# \mathcal{D}_{idi}} {Time},
\end{equation}}}
where $Time$ represents the consumed time during the generation procedure. Thus, a larger value of $ \# \mathcal{D}_{idi}$ means the method is more effective, and a higher value of $EGS$ means the method is more efficient. 

\textbf{Naturalness Evaluation}. Naturalness measures how similar the distribution of generated instances is compared with the original dataset. 
\revised{To thoroughly evaluate the naturalness, we adopt a comprehensive metric similarly in \cite{wen2021causal}, which} considers both the distribution of each column and the correlation between pairs of columns in evaluation. 

As for the distribution, we use Kolmogorov-Smirnov Test (\revised{for numerical column}) and the Total Variation Distance (\revised{for categorical column}) to calculate its score. As for the correlation, we use Pearson Coefficient (\revised{for two numerical columns}) and the Contingency Similarity (\revised{for categorical column with any kind of column}) to calculate its score. And we compute the average of these two scores to measure the naturalness of $ \mathcal{D}_{idi}$, denoted as $ATN$ (Average Tabular Naturalness): 
\revised{
\small{
\begin{equation} 
\begin{aligned}
\label{equation:ATN}
ATN(\mathcal{D}_{idi},X) = & \frac{1}{2} [\rm{mean}(
\{  \rm{KS}(\mathcal{D}_{{idi}_{p}},X_{p}) 
: \emph{X}_{p} \, \rm{is} \, \rm{numerical} 
\}    \\
& \cup \{ \rm{TV}(\mathcal{D}_{{idi}_{p}},\emph{X}_{p}) 
: \emph{X}_{p} \, \rm{is} \, \rm{categorical}  
\} ) \, +  \\
&  \rm{mean}(
\{  \rm{PC}(\mathcal{D}_{{idi}_{p,q}},\emph{X}_{p,q}) 
: \emph{X}_{p}, \emph{X}_{q} \, \rm{are} \, \rm{numerical} 
\} \\
&  \cup \{ \rm{CS}(\mathcal{D}_{{idi}_{p,q}},\emph{X}_{p,q}) 
: \emph{X}_{p} \, \rm{is} \, \rm{categorical}  
\} )
]
\end{aligned}
\end{equation}
}}
\revised{where $\rm{mean}(\cdot)$ calculates the average value of set,  $X_{p}$ denotes the pth column of $X$, and KS is an inverted version of Kolmogorov-Smirnov Test. }
The range of $ATN$ is 0 to 1, and we say that a method presents a better naturalness if its generated $ \mathcal{D}_{idi}$ achieves a higher value of $ATN$. We use the implementation provided by SDMetrics \cite{sdmetrics} library for evaluation. \revised{Besides $ATN$, we also evaluate naturalness utilizing both classifier-based detection (LogisticRegression, SVC) and distance-based method (Average Nearest Neighbor Euclidean distances). Their details can be found on our website \cite{ourweb}.}

\textbf{Fairness evaluation}. Both individual fairness and group fairness are taken into evaluation. 
To measure individual fairness, we first follow previous work  \cite{udeshi2018automated} to randomly sample a large set of instances and check the ratio of discriminatory instances in the set (denoted as $IF_{r}$); moreover, we calculate the proportion of individual discriminatory instances that exists in the original dataset (denoted as $IF_{o}$):
\revised{\small{
\begin{equation} 
IF_{r} = \frac{\# \mathcal{D}_{idi}}{\vert \rm{randomly \, sampled \, set} \vert}, \, IF_{o} = \frac{\# \mathcal{D}_{idi}}{\vert \rm{original \, dataset} \vert}.
\end{equation}
}}

To measure group fairness, we adopt \revised{Statistical Parity Difference (SPD)} and \revised{Average Odds Difference (AOD)} metrics following \cite{hort2021fairea}. The SPD requires that a decision should be independent of the protected attributes \cite{barocas2016big}, which measures the difference in positive classification between different demographic groups:
\revised{\small{
\begin{equation} 
SPD = \vert P(\hat{y}=1|A=unprivileged)-P(\hat{y}=1|A=privileged) \vert,
\end{equation}
where $\hat{y}$ denotes the model prediction and $A$ is the group of the protected attribute. 
}}

The AOD represents the average of the differences in True Positive Rate (TPR) and False Positive Rate (FPR) between privileged and unprivileged groups \cite{hardt2016equality}:
\revised{\small{
\begin{equation}
\begin{aligned}
AOD = & \frac{1} {2}[ \vert FPR_{A=unprivileged}- FPR_{A=privileged} \vert  \\ 
& +  \vert TPR_{A=unprivileged}- TPR_{A=privileged} \vert].
\end{aligned}
\end{equation}
 }}
 
Under the above definitions, lower fairness metric values indicate a fairer model, while larger values denote a higher level of discrimination in the model.
\vspace{-0.05in}
\subsubsection{Implementation details} We adopt the CTGAN \cite{xu2019modeling} as the GAN model in experiments. We train the model with 300 epochs and the batch size is set as 500 for each tabular dataset. The average consumed training time is 13 minutes. 
For the surrogate boundary, we implement it with a linear SVM trained on \revised{the refined randomly sampled latent dataset $\mathcal{D}_{aux}$ (containing 100K latent vectors).}
Without specification, we set the hyperparameter $\lambda$ as 0.3. We conduct our experiments on a server with Intel(R) Xeon(R) Gold 6230R CPU @ 2.10GHz, 256GB system memory, and an NVIDIA GeForce RTX-2080-Ti GPU.

\vspace{-0.1in}
\subsection{RQ1: Effectiveness and Efficiency}

For the baselines, we \revised{follow their original two-stage settings \cite{aggarwal2019black,zhang2020white,zhang2021efficient,zheng2022neuronfair,fan2022explanation,udeshi2018automated}, which} first search 1000 instances to construct individual discriminatory seeds in the global phase and then search 1000 neighbor instances for each seed (the maximum number of total test cases is 1K $\times$ 1K = 1M). 
As for our approach, we use the entire randomly sampled dataset $\mathbf{Z}_{init}$ (contains 1M latent vectors) as start points and set the maximum number of test cases as 1M for fair comparisons. 
\revised{We opt to maintain the original two-stage framework of the baselines since that directly combining the two search numbers and restricting the total number of test cases to 1M would disrupt the framework and adversely affect their performance.}
Following \cite{fan2022explanation}, we constrain the testing time as one hour. To mitigate contingency, we repeat the generation process 5 times and report the average results.

The results on DNN and RF are shown in Table \ref{tab:main-attack} and \ref{tab:rf-attack}, while the results on SVM testing can be found on our website \cite{ourweb}. Note that the gradient-based methods (\ie, ADF, EIFIG, and NeuronFair) are unavailable for the traditional machine learning models, and we only apply them to DNN models. From the results, we could make several \textbf{observations} as follows:

\begin{itemize}[leftmargin=*]
\item \emph{As for the generation effectiveness,} our approach generates more individual discriminatory instances than other baselines and outperforms them largely (+135\revised{,}534 on average). Specifically, for DNN models, as shown in Table \ref{tab:main-attack}, the average $ \# \mathcal{D}_{idi}$ on four datasets of \tool is \textbf{149\revised{,}711}, which is \textbf{18 times} and \textbf{3 times} better than NeuronFair and ExpGA, respectively; for RF models, as shown in Table \ref{tab:rf-attack}, our method finds \textbf{2.5 times} and \textbf{2.2 times} more discriminatory instances than SG and ExpGA; for SVM models, we also achieve good performance, especially on the Credit and Meps datasets. The above results demonstrate the outstanding performance of our method on the test instances generation effectiveness, which can be attributed to our surrogate decision boundary approximation and candidate probing strategies.

\item \emph{As for the generation efficiency}, our \tool has a faster generation speed and outperforms other baselines by 8.71 times on average. For example, as shown in Table \ref{tab:main-attack}, \tool can generate almost \textbf{50 discriminatory instances per second} on average while the value of NueronFair and ExpGA is 2 and 24, respectively. This indicates that our \tool testing method is capable to give rapid feedback to software engineers, which will be preferable in time-constrained industrial software testing. We speculate the main reason is the direct vector calculation employed in the latent candidates probing phase, which can reduce the computational complexity and circumvent the tedious iterative search process.

\item Moreover, we observe that our method shows \emph{better stability across different combinations of datasets and target models}. Specifically, for ExpGA, when testing the RF model on the Adult dataset with race as the protected attribute, $ \# \mathcal{D}_{idi}$ is only 286, which is much lower than its performance in other test conditions (\eg, $ \# \mathcal{D}_{idi}$ is 10\revised{,}283 on the Adult with gender as the protected attribute); for AEQUITAS and SG, when testing SVM models on the Adult dataset, $ \# \mathcal{D}_{idi}$ even reduces to 0, denoting a test failure happened. In fact, both SG and ExpGA highly depend on the explanation results produced by the local explainer, which constrains their generation when the interpretation is weak. By contrast, our \tool only requires the prediction results of the tested model and learns the surrogate decision boundary from the model decision to guide testing, which shows better tests for different models/datasets combinations.

\end{itemize}
\vspace{-0.05in}
\begin{tcolorbox}[size=title]
	{\textbf{Answer to RQ1:} In summary, \tool outperforms 6 baselines significantly in both effectiveness and efficiency. On average, \tool generates \textbf{$\times$9.42} more individual discriminatory instances at \textbf{$\times$8.71} faster speed.}
\end{tcolorbox}
\vspace{-0.2in}

\subsection{RQ2: Naturalness of $\mathcal{D}_{idi}$}
\vspace{-0.05in}

\begin{table*}[ht]
\caption{The effectiveness and efficiency of different generation methods on different datasets against DNN models.}
\vspace{-0.15in}
  \label{tab:main-attack}
  \resizebox{\linewidth}{!}{
\begin{tabular}{@{}ll|rr|rr|rr|rr|rr|rr|rr@{}}
\toprule
\multirow{2}{*}{Dataset} & Pro. & \multicolumn{2}{c|}{AEQUITAS}                     & \multicolumn{2}{c|}{SG}  & \multicolumn{2}{c|}{ADF} & \multicolumn{2}{c|}{EIDIG}  & \multicolumn{2}{c|}{NeuronFair} & \multicolumn{2}{c|}{ExpGA}  & \multicolumn{2}{c}{\tool}  \\ \cmidrule(l){3-16} 
&
  Att. & $\#\mathcal{D}_{idi}$$\textcolor{red}{\uparrow}$ & \multicolumn{1}{c|}{$EGS$$\textcolor{red}{\uparrow}$} & $\#\mathcal{D}_{idi}$$\textcolor{red}{\uparrow}$ & \multicolumn{1}{c|}{$EGS$$\textcolor{red}{\uparrow}$} & $\#\mathcal{D}_{idi}$$\textcolor{red}{\uparrow}$ & \multicolumn{1}{c|}{$EGS$$\textcolor{red}{\uparrow}$} & $\#\mathcal{D}_{idi}$$\textcolor{red}{\uparrow}$ & \multicolumn{1}{c|}{$EGS$$\textcolor{red}{\uparrow}$} & $\#\mathcal{D}_{idi}$$\textcolor{red}{\uparrow}$ & \multicolumn{1}{c|}{$EGS$$\textcolor{red}{\uparrow}$} & $\#\mathcal{D}_{idi}$$\textcolor{red}{\uparrow}$ & \multicolumn{1}{c|}{$EGS$$\textcolor{red}{\uparrow}$} & $\#\mathcal{D}_{idi}$$\textcolor{red}{\uparrow}$ & $EGS$$\textcolor{red}{\uparrow}$ \\ \midrule
\multirow{3}{*}{Adult}     & gender & 3,484 &  1.09  & 1,806  & 0.50 &  11,540 & 3.20 & 12,503 &  3.47 & 12,791 & 3.55 & 40,877 & 39.08  &  \pmb{86,065} & \pmb{41.71}   \\
                           & race   & 3,393 &  0.94  & 2,893 & 0.80  & 11,827 &  3.29 &  11,330 & 3.15 & 11,991 & 3.33 & 4,215 & 9.73 & \pmb{132,018}  &  \pmb{36.67}      \\
                           & age    & 3,982 &  1.11  & 148,58 & 4.13  & 12,078 &  3.36 & 12,382 & 3.44 &  13,214 & 3.67 & 45,615 & 12.61 & \pmb{148,221}  &   \pmb{41.17}     \\ \midrule
\multirow{2}{*}{Credit}    & gender & 1,671 &  0.53  & 9,241 & 2.70  & 4,842 &  1.35 &   5,201 & 1.44 &  4,469 & 1.24 & 158,687 & 44.05 & \pmb{164,569}  &  \pmb{69.38}       \\
                           & age   & 1,645 &  0.46  & 12,854 & 3.57  & 6,090 &  1.69 &   6,494 & 1.80 &  6,532 & 1.81 & 43,202 & 11.98 & \pmb{261,504}  &  \pmb{72.64}    \\ \midrule
Bank                      & age   & 1,435 &  0.40  & 9,276 & 2.58   &  3,979 &  1.11 &   4,731 & 1.31 &  5,143 & 1.43 & 24,691 & 6.84 & \pmb{145,901}  &  \pmb{40.53}    \\ \midrule
Meps                      & gender  & 1,549 &  0.43  & 5,922 &  1.65 & 5,555 &  1.54 &   4,914 & 1.37 &  3,970 & 1.10 & 32,580 & 46.78 & \pmb{109,701}  & \pmb{48.67}  \\ \bottomrule
\end{tabular}
}
\vspace{-0.175in}
\end{table*}

\begin{table}[ht]

\caption{The effectiveness and efficiency of different generation methods on different datasets against RF models.}
\vspace{-0.15in}
  \label{tab:rf-attack}
\resizebox{\columnwidth}{!}{%
\begin{tabular}{@{}ll|rr|rr|rr|rr@{}}
\toprule
\multirow{2}{*}{Dataset} & Pro.   & \multicolumn{2}{c|}{AEQUITAS} & \multicolumn{2}{c|}{SG} & \multicolumn{2}{c|}{ExpGA} & \multicolumn{2}{c}{\tool} \\ \cmidrule(l){3-10} 
 &
  Att. &
  $\#\mathcal{D}_{idi}$$\textcolor{red}{\uparrow}$ &
  $EGS$$\textcolor{red}{\uparrow}$ &
  $\#\mathcal{D}_{idi}$$\textcolor{red}{\uparrow}$ &
  $EGS$$\textcolor{red}{\uparrow}$ &
  $\#\mathcal{D}_{idi}$$\textcolor{red}{\uparrow}$ &
  $EGS$$\textcolor{red}{\uparrow}$ &
  $\#\mathcal{D}_{idi}$$\textcolor{red}{\uparrow}$ &
  $EGS$$\textcolor{red}{\uparrow}$ \\ \midrule
\multirow{3}{*}{Adult}   & gender & 441 & 0.12 & 2,487 & 0.69 & 10,283 & 2.85 & \pmb{13,360} & \pmb{3.71}      \\
     & race   & 100 & 0.03 & 1,316 & 0.37 & 286 & 0.08 & \pmb{10,398} & \pmb{2.89}\\
     & age    & 846 & 0.24 & 5,152 & 1.43 & 9,402 & 2.61 & \pmb{14,185} & \pmb{3.94}\\ \midrule
\multirow{2}{*}{Credit}  & gender & 219 & 0.06 & 7,338 & 2.04 & 1,799 & 0.50 & \pmb{15,010} & \pmb{4.17}\\
     & age    & 153 & 0.04 & 7,908 & 2.20 & 946 & 0.26 & \pmb{11,972} & \pmb{3.33}\\ \midrule
Bank & age    & 186 & 0.05 & 6,104 & 1.70 & 4,071 & 1.13 & \pmb{6,578} & \pmb{1.83}\\ \midrule
Meps & gender & 276 & 0.08 & 3,865 & 1.07 & 3,472 & 0.96 & \pmb{3,958} & \pmb{1.10}\\ \bottomrule
\end{tabular}%
}
\vspace{-0.2in}
\end{table}

\begin{table}[ht]
\caption{The naturalness of discriminatory instances generated for DNN models. Results are shown in $ATN$.}
\vspace{-0.15in}
  \label{tab:naturalness-dnn}
\resizebox{\columnwidth}{!}{%
\begin{tabular}{@{}l|rrr|rr|r|r@{}}
\toprule
\multirow{2}{*}{Methods} & \multicolumn{3}{c|}{Adult}                          & \multicolumn{2}{c|}{Credit}       & \multicolumn{1}{c|}{Bank} & \multicolumn{1}{c}{Meps} \\ \cmidrule(l){2-8} 
           & gender & race   & age    & gender & age    & age    & gender \\ \midrule
AEQUITAS  & 50.53\% & 48.02\% & 46.22\% & 67.84\% & 65.02\% & 60.85\% & 69.45\% \\
SG      & 43.23\% & 46.10\% & 46.03\% & 55.45\% & 54.49\% & 41.78\% & 53.03\% \\
ADF & 63.79\% & 66.91\% & 64.24\% & 73.13\% & 74.37\% & 71.14\% & 70.17\% \\
EIDIG     & 63.45\% & 66.73\% & 64.69\% & 73.67\% & 73.97\% & 69.62\% & 71.09\% \\
NeuronFair      & 63.93\% & 66.89\% & 64.33\% & 73.97\% & 74.18\% & 71.56\% & 71.11\% \\  
ExpGA      & 41.20\% & 48.76\% & 50.23\% & 60.53\% & 63.30\% & 58.54\% & 64.83\% \\
\tool   & \pmb{80.98\%} & \pmb{81.94\%} & \pmb{82.10\%} & \pmb{80.15\%} & \pmb{80.05\%} & \pmb{80.84\%}   & \pmb{82.93\%}   \\ 
\bottomrule
\end{tabular}%
}
\vspace{-0.2in}
\end{table}

To answer this question, we randomly select the same number of instances as the original dataset from our constructed $\mathcal{D}_{idi}$ and then measure $ATN$ for each sampled set. To ensure sampling adequacy, we repeat 10 times and report the average $ATN$ values. Considering the $ \# \mathcal{D}_{idi}$ values of some methods are less than the quantity of the original dataset, we remove the time limit and obtain sufficient discriminatory instances for testing.

From the results in Table \ref{tab:naturalness-dnn}, we identify that the $ATN$ values of our method are significantly higher than that of other baselines on all datasets. Specifically, the $ATN$ values of our method in all settings are higher than 80\%, indicating that the generated instances by \tool are more natural. \emph{In other words, the distribution of generated instances is much closer to the real data distribution.} This observation could be used to further explain the outstanding performance of \tool as it generates test cases near the real decision boundary of the target model that reflects the nature of the original data distribution. 
In more detail, the white-box methods (\ie, ADF, EIDIG, and NeuronFair) utilize gradients to perturb instances to the decision boundary and show \revised{second} naturalness (69.11\%, 69.03\%, and 69.42\% on average on all datasets). 
\revised{
    The impaired naturalness in these white-box methods can be attributed to modifying directly on the raw input (\ie, attribute value) without any restriction, unlike \tool which probes instances in the semantic latent space implicitly limited by the GAN. This reason has also been observed in image data \cite{riccio2020model}.
}
As for black-box methods (\ie, AEQUITAS, SG, and ExpGA), they show the worst naturalness (58.28\%, 48.59\%, and 55.34\% on average on all datasets). We speculate that the randomness in AEQUITAS and SG, as well as the crossover and mutation operator employed in ExpGA, induce the worst naturalness. \revised{As for the classifier-based and distance-based measures, our \tool still outperforms other baselines, and we also observe consistent trends in naturalness (\tool > white-box methods > black-box methods). Further details are reported on our website \cite{ourweb}.}

\begin{figure}[ht]
\vspace{-0.04in}
    \centering
    \subfigure[Adult-gender]{
    \vspace{-0.15in}
        \includegraphics[width=0.47\linewidth]{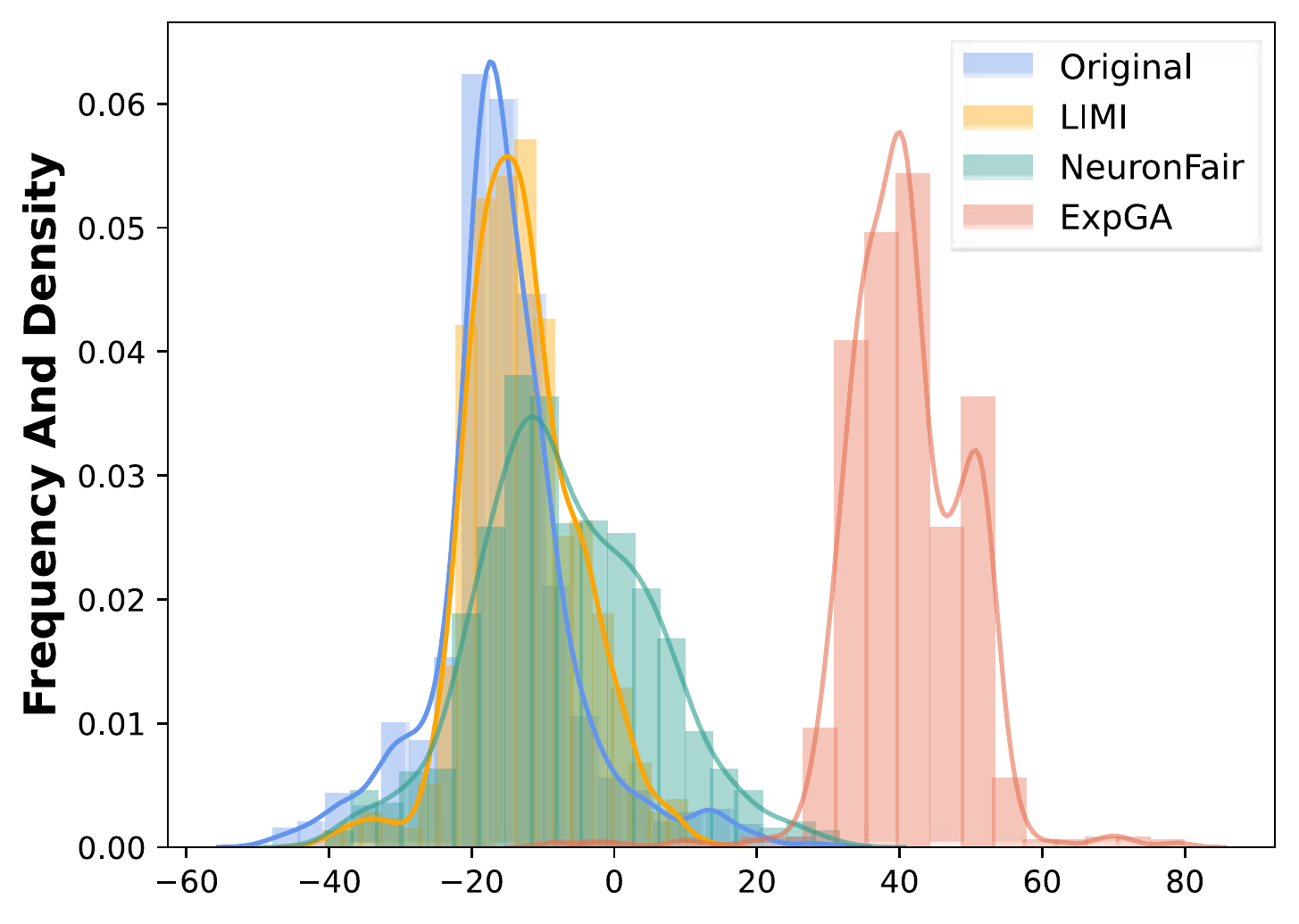}
    }
    \subfigure[Adult-race]{
    \vspace{-0.15in}
        \includegraphics[width=0.47\linewidth]{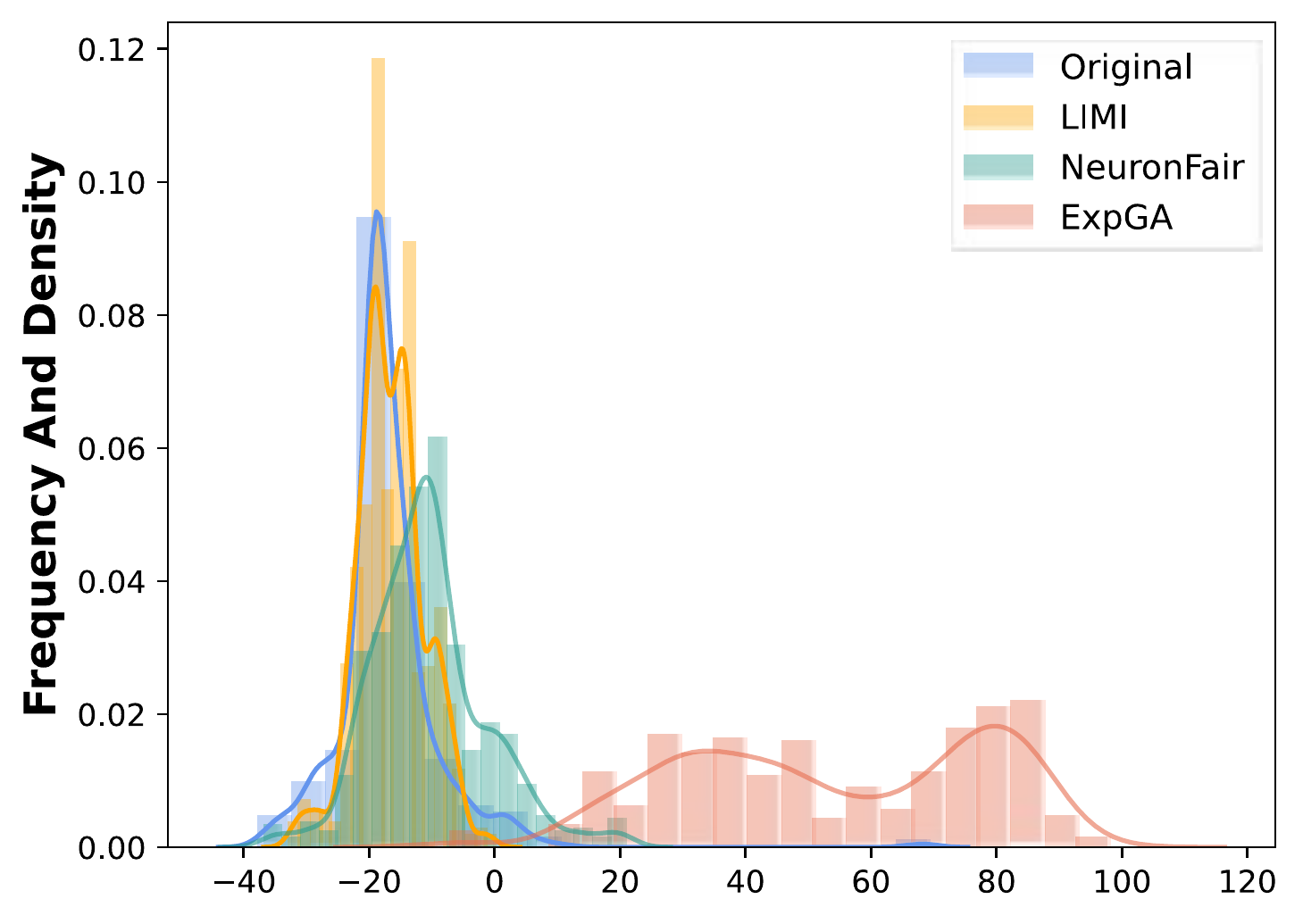}
    }
    
    \vspace{-0.2in}
    \caption{Visualization of the data distribution of generated instances. The distribution of our \tool is the closest to the original one, which indicates the better naturalness of generated instances.}
    \vspace{-0.2in}
    \label{fig:freq}
\end{figure}

To better illustrate the naturalness, we visualize the feature similarity between the generated and original data for NeuronFair, ExpGA, and our \toolns. In particular, we randomly select 1000 discriminatory instances on the DNN models and adopt PCA  \cite{daffertshofer2004pca} to translate their features into one dimension; we then report and visualize the feature distribution of these instances. As shown in Figure \ref{fig:freq}, we can identify that the distribution of instances generated by \tool (shown in orange) is the closest to the original data distribution, which indicates the better naturalness of our \tool generation and further verifies our motivation.

\vspace{-0.05in}
\begin{tcolorbox}[size=title]
	{\textbf{Answer to RQ2:} The individual discriminatory instances generated by \tool \revised{possess} better naturalness (more similar to the original data distribution), showing an average \textbf{19.65\%} improvement in terms of $ATN$.}

\end{tcolorbox}
\vspace{-0.05in}

\vspace{-0.1in}
\subsection{RQ3: Fairness Improvement}
\vspace{-0.05in}

\begin{table*}[ht]
\caption{The fairness improvement experiments via retraining DNN models.}
\vspace{-0.15in}
  \label{tab:fair-dnn}
\resizebox{\linewidth}{!}{%
\begin{tabular}{@{}ll|rrrrr|rrrrr|rrrrr|rrrrr|rrrrr|rrrrr@{}}
\toprule
\multirow{2}{*}{Dataset} &
  \multirow{2}{*}{Pro.Attr.} &
  \multicolumn{5}{c|}{Before} &
  \multicolumn{5}{c|}{\revised{SG}} &
  \multicolumn{5}{c|}{\revised{ADF}} &
  \multicolumn{5}{c|}{NeuronFair} &
  \multicolumn{5}{c|}{ExpGA} &
  \multicolumn{5}{c}{\tool} \\ \cmidrule(l){3-32} 
                        &      & ACC$\textcolor{red}{\uparrow}$ & $IF_r$$\textcolor{green}{\downarrow}$ & $IF_o$$\textcolor{green}{\downarrow}$ & $SPD$$\textcolor{green}{\downarrow}$ & $AOD$$\textcolor{green}{\downarrow}$ 
                        & ACC$\textcolor{red}{\uparrow}$ & $IF_r$$\textcolor{green}{\downarrow}$ & $IF_o$$\textcolor{green}{\downarrow}$ & $SPD$$\textcolor{green}{\downarrow}$ & $AOD$$\textcolor{green}{\downarrow}$ 
                        & ACC$\textcolor{red}{\uparrow}$ & $IF_r$$\textcolor{green}{\downarrow}$ & $IF_o$$\textcolor{green}{\downarrow}$ & $SPD$$\textcolor{green}{\downarrow}$ & $AOD$$\textcolor{green}{\downarrow}$ 
                        & ACC$\textcolor{red}{\uparrow}$ & $IF_r$$\textcolor{green}{\downarrow}$ & $IF_o$$\textcolor{green}{\downarrow}$ & $SPD$$\textcolor{green}{\downarrow}$ & $AOD$$\textcolor{green}{\downarrow}$ 
                        & ACC$\textcolor{red}{\uparrow}$ & $IF_r$$\textcolor{green}{\downarrow}$ & $IF_o$$\textcolor{green}{\downarrow}$ & $SPD$$\textcolor{green}{\downarrow}$ & $AOD$$\textcolor{green}{\downarrow}$ 
                        & ACC$\textcolor{red}{\uparrow}$ & $IF_r$$\textcolor{green}{\downarrow}$ & $IF_o$$\textcolor{green}{\downarrow}$ & $SPD$$\textcolor{green}{\downarrow}$ & $AOD$$\textcolor{green}{\downarrow}$  \\ \midrule
 
\multirow{3}{*}{Adult}  & gender & 88.16\% & 3.59\% & 8.20\% & 0.17 & 0.04 & \revised{87.94\%} & \revised{3.45\%}  & \revised{7.72\%}  & \revised{0.16} & \revised{0.04} & \revised{87.51\%} & \revised{3.02\%}  & \revised{7.52\%}  & \revised{0.17} & \revised{0.04}  & 87.62\% & 3.18\% & 7.03\% & 0.16 & 0.04  & 88.44\% & 3.34\% & 8.23\% & 0.16 & \pmb{0.03}  & 86.92\% & \pmb{1.67\%} & \pmb{5.50\%} & \pmb{0.14} & \pmb{0.03}   \\
                        & race  & 88.16\% & 7.13\% & 16.31\% & 0.09 & 0.04 & \revised{88.21\%} & \revised{6.33\%}  & \revised{14.05\%} & \revised{\pmb{0.08}} & \revised{0.04}  & \revised{87.66\%} & \revised{6.45\%}  & \revised{13.56\%} & \revised{\pmb{0.08}} & \revised{0.04}
                        & 87.42\% & 6.15\% & 13.13\% & 0.09 & 0.04 & 88.33\% & 6.94\% & 15.95\% & 0.09 & 0.06  & 87.04\% & \pmb{2.51\%} & \pmb{8.63\%} & \pmb{0.08} & \pmb{0.02}  \\
                        & age   & 88.16\% & 10.72\% & 26.92\% & 0.21 & 0.19 & \revised{88.18\%} & \revised{9.10\%}  & \revised{24.70\%} & \revised{0.20} & \revised{0.10}  & \revised{87.78\%} & \revised{8.95\%}  & \revised{25.64\%} & \revised{0.21} & \revised{0.11}  & 87.73\% & 7.78\% & 25.49\% & 0.22 & 0.12  & 88.34\% & 10.30\% & 25.47\% & 0.20 & 0.11  & 86.86\% & \pmb{5.91\%} & \pmb{17.34\%} & \pmb{0.17} & \pmb{0.05} \\ \midrule
\multirow{2}{*}{Credit} & gender & 100\% & 14.09\% & 15.50\% & \pmb{0.07} & \pmb{0.00} & \revised{99.17\%} & \revised{15.76\%} & \revised{16.00\%} & \revised{\pmb{0.07}} & \revised{\pmb{0.00}} & \revised{99.23\%} & \revised{12.11\%} & \revised{10.33\%} & \revised{\pmb{0.07}} & \revised{\pmb{0.00}}  & 99.20\% & 11.65\% & 10.87\%& \pmb{0.07} & \pmb{0.00}  & 99.90\% & 14.04\% & 11.20\% & \pmb{0.07} & \pmb{0.00}  & 99.83\% & \pmb{11.18\%}  & \pmb{9.30\%}  & \pmb{0.07} & \pmb{0.00} \\
                       & age   & 100\% & 41.96\% & 42.17\% & \pmb{0.00} & \pmb{0.00} & \revised{99.20\%} & \revised{36.40\%} & \revised{37.20\%} & \revised{0.01} & \revised{0.01}  & \revised{99.80\%} & \revised{36.74\%} & \revised{34.90\%} & \revised{\pmb{0.00}} & \revised{\pmb{0.00}}  & 99.97\% & 36.61\% & 35.67\% & \pmb{0.00} & \pmb{0.00}  & 100\% & 37.82\% & 38.20\% & \pmb{0.00} & \pmb{0.00}  & 99.97\% & \pmb{29.57\%} & \pmb{34.60\%} & \pmb{0.00} & \pmb{0.00}  \\ \midrule
Bank                    & age   & 93.65\% & 25.79\% & 27.94\% & 0.23 & \pmb{0.07} & \revised{93.37\%} & \revised{25.83\%} & \revised{27.33\%} & \revised{0.29} & \revised{0.13} & \revised{93.34\%} & \revised{24.57\%} & \revised{27.13\%} & \revised{0.24} & \revised{0.08}  & 93.28\% & 24.15\% & 26.38\% & 0.23 & \pmb{0.07}  & 93.10\% & 24.38\% & 25.73\% & 0.28 & 0.12 & 92.50\% & \pmb{13.15\%} & \pmb{16.98\%} & \pmb{0.21} & \pmb{0.07}  \\ \midrule
Meps                    & gender & 95.07\% & 7.78\% & 11.72\% & 0.08 & 0.02 & \revised{94.13\%} & \revised{8.44\%}  & \revised{10.85\%} & \revised{\pmb{0.07}} & \revised{\pmb{0.01}}  & \revised{93.98\%} & \revised{6.86\%}  & \revised{11.46\%} & \revised{0.08} & \revised{\pmb{0.01}}  & 94.13\% & 6.70\% & 10.14\% & \pmb{0.07} & 0.02  & 93.05\% & 7.02\% & 10.83\% & \pmb{0.07} & \pmb{0.01}  & 94.77\% & \pmb{3.32\%} & \pmb{9.74\%} & \pmb{0.07} & \pmb{0.01}    \\ \bottomrule

\end{tabular}%
}
\vspace{-0.2in}
\end{table*}

To further demonstrate the importance of the naturalness of the generated discriminatory instances, we follow \cite{zhang2020white} and retrain the models using the generated instances to improve model fairness. Given the number of generated instances is much larger than the original dataset, we randomly select the instances with 30\% size of the original dataset for retraining, which is almost the same quantity to \cite{zhang2020white}. We use the selected instances and the original dataset to retrain DNN models and use the fairness metrics (\ie, $IF_r$, $IF_o$, $SPD$, and $AOD$) to evaluate the fairness improvement. \revised{We also evaluate the performance of retraining RF models where we achieve similar results to DNNs. Due to the space limitation, the results can be found on our website \cite{ourweb}.} For fair comparisons, we repeat the procedure 5 times and report the average values to avoid the effect of randomness.

The results regarding both the individual and group fairness are shown in Table \ref{tab:fair-dnn}, where column ‘Before’ denotes the model trained on the original dataset, and columns named by the approaches denote the model retrained with discriminatory instances they generated. From the results, we have the following \textbf{observations:}

\begin{itemize}[leftmargin=*]
\item \emph{As for the accuracy}, the retrained models have a slight degradation (less than 1\% on average), indicating that the models retrained with discriminatory instances can still perform well in real-world datasets.   

\item \emph{As for the individual fairness}, our \tool outperforms other baselines significantly in terms of both $IF_r$ and $IF_o$. On average, \tool achieves individual fairness improvement of \textbf{45.67\%} in terms of $IF_r$, and \textbf{32.81\%} in terms of $IF_o$. In more detail, the models retrained by our method achieve \textbf{3.11 times} and \textbf{2.22 times} than NeuronFair, and \textbf{8.20 times} and \textbf{3.84 times} than ExpGA on individual fairness under two metrics. The above results indicate that individual discriminatory instances generated by our method show better effectiveness in the model fairness improvement.

\item \emph{As for the group fairness}, we also notice that retrained models enjoy improvements on almost all datasets. 
However, the improvements are comparatively slight (+0.02 in terms of $SPD$, and +0.03 in terms of $AOD$ on average). 
For instance, on the Adult dataset, \tool achieves \textbf{15.94\%} and \textbf{49.56\%} improvements of $SPD$ and $AOD$ respectively; by contrast, retraining by NeuronFair and ExpGA even make the models more unfair between different groups on the Adult and Bank datasets. We speculate that the main reason is that the distribution of instances generated by \tool is closer to the original dataset, while instances generated by NeuronFair and ExpGA deviate from the real distribution. We also observe that for small dataset Credit, the model overfits on the dataset and behaves even with no discrimination. Thus, directly retraining models using new discriminatory instances would not increase fairness on the Credit. We will further study it in the future.

\end{itemize}
\vspace{-0.05in}
\begin{tcolorbox}[size=title]
	{\textbf{Answer to RQ3:} The natural discriminatory instances generated by \tool are useful for fairness improvement. On average, we achieve individual fairness improvements of \textbf{45.67\%} on $IF_r$ and \textbf{32.81\%} on $IF_o$, and group fairness improvements of \textbf{9.86\%} on $SPD$ and \textbf{28.38\%} on $AOD$.}

\end{tcolorbox}
\vspace{-0.05in}

\vspace{-0.1in}
\section{Investigations and Analyses} \label{sec:discussion}
\vspace{-0.05in}

Besides the main experiments conducted in the previous section, this section conducts further investigations and analyses to better understand our proposed framework in terms of fitness, stability, scalability, and generalizability.

\begin{table}[ht]
  \caption{The fitness of surrogate decision boundaries. Results are shown in AUC.}
\vspace{-0.175in}
  \label{tab:fitness}
\begin{tabular}{@{}l|rr|rr|rr@{}}
\toprule
\multirow{2}{*}{Dataset} & \multicolumn{2}{c|}{DNN} & \multicolumn{2}{c|}{RF}  & \multicolumn{2}{c}{SVM} \\ \cmidrule(l){2-7} 
                         & Train & Entire & Train & Entire & Train & Entire   \\ \midrule
Adult   & 96.71\% & 86.43\% & 99.78\% & 89.72\% & 99.98\% & 93.94\% \\ \midrule
Credit  & 82.16\% & 67.82\% & 96.69\% & 76.54\% & 99.93\% & 75.36\% \\ \midrule
Bank  & 95.47\% & 85.61\% & 99.69\% & 94.59\% & 99.65\% & 96.40\% \\ \midrule
Meps  & 93.52\% & 84.09\% & 99.96\% & 97.20\% & 99.42\% & 90.50\% \\ \bottomrule
\end{tabular}%
\vspace{-0.175in}
\end{table}

\begin{table}[ht]
\caption{Stability study. Ablations on the influence of $\lambda$ using DNN models. Results are shown in $\# \mathcal{D}_{idi}$.}
\vspace{-0.15in}
  \label{tab:lambda-attack}
\resizebox{\columnwidth}{!}{%
\begin{tabular}{@{}llrrrrrr@{}}
\toprule
Dataset                 & Pro.Attr. & $\lambda=0$ & $\lambda=0.1$  & $\lambda=0.2$ & $\lambda=0.3$ & $\lambda=0.4$ & $\lambda=0.5$ \\ \midrule

\multirow{3}{*}{Adult}  & gender & 38,974 & 80,947  & \pmb{89,348}  & 86,065  & 78,188  & 69,989  \\
                        & race   &  76,294      & 100,587 & 109,761 & \pmb{132,018} & 109,826 & 114,051 \\
                        & age    & 54,644      & 99,365  & 110,158 & \pmb{148,221} & 115,063 & 110,951 \\ \midrule
\multirow{2}{*}{Credit} & gender & 90,210      & \pmb{168,876} & 164,561 & 164,569 & 165,417 & 165,550 \\
                        & age    &  152,493     & 204,704 & 196,113 & \pmb{261,504} & 196,799 & 201,052 \\ \midrule
Bank                    & age    &  50,820     & 77,396  & 94,739  & \pmb{145,901} & 116,837 & 118,272 \\ \midrule
Meps                    & gender &  49,769     & \pmb{130,152} & 128,876 & 109,701 & 88,500  & 73,054 \\
\bottomrule
\end{tabular}%
}
\vspace{-0.225in}
\end{table}

\textbf{Fitness}. We first evaluate the fitness of our surrogate decision boundary to the target model through the Area Under Receiver Operating Characteristic Curve (AUC) metric \cite{hanley1982meaning}, which measures the fitness in the class-imbalanced latent samples. Specifically, we evaluate our surrogate boundaries (\ie, linear SVM) both on the training set (100K latent samples with high classification confidence) and the entire set (1M random latent samples). The fitness results are shown in Table \ref{tab:fitness}. Specifically, on Adult, Bank, and Meps datasets, our surrogate boundaries achieve over 93\% AUC on the high-confidence training set, and over 84\% AUC on the imbalanced entire set, which shows a proper approximation of the real decision boundary of the tested model. However, the AUC value is relatively lower on the Credit dataset, especially on the DNN model (67.82\% for the entire latent samples). We attribute this observation to the small size of the Credit dataset (\ie, only 600 samples), which would cause an over-fitted DNN that fails to generalize well to a large number of new samples.

\textbf{Stability}. During the potential probing process, the walking stepsize $\lambda$ along the unit vector is critical for finding the candidates. Thus, we hereby investigate the influence of hyperparameter $\lambda$ on extra ablation studies. Specifically, we use our \tool to test a DNN model within an hour and set $\lambda$ as 0, 0.1, 0.2, 0.3, 0.4, and 0.5, respectively. As shown in Table \ref{tab:lambda-attack}, the performance of \tool is relatively stable with only slight fluctuations for different $\lambda$ values. Specifically, we observe that $\lambda=0.3$ generate \textbf{2} times more instances than $\lambda=0$, indicating the effectiveness of our candidate probing strategy; in the worst case setting ($\lambda=0.5$), $\# \mathcal{D}_{idi}$ drops 18.61\% on average, however, it still achieves comparable performance to other baselines (\emph{c.f.} Table \ref{tab:main-attack}).  
Thus, we set $\lambda=0.3$ in our main experiments.
Overall, our \tool behaves stably when the hyperparameter $\lambda$ falls in the range of $[0.1,0.5]$.
\revised{Ulteriorly, we analyze the influence of latent boundary approximation, where we replace the approximated surrogate boundary with a randomly generated boundary. The average $\# \mathcal{D}_{idi}$ value (56,637 for approximated boundary vs 8,418 for randomly generated boundary) on the Adult dataset, demonstrating the effectiveness of our boundary approximation module.
}

\begin{figure}[ht]
\vspace{-0.1in}
    \centering
    \includegraphics[width=\linewidth]{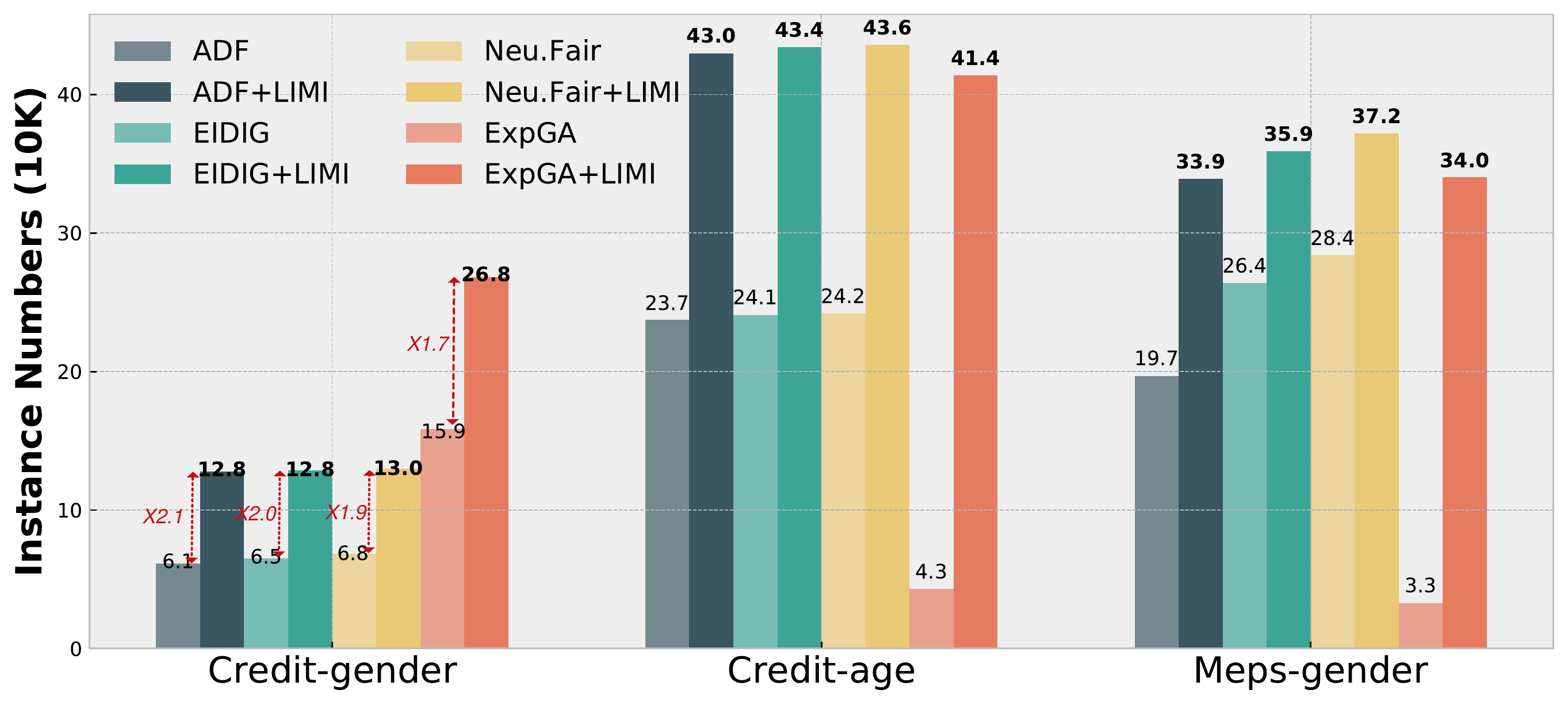}
    \vspace{-0.325in}
    \caption{Scalability study. Comparison between baselines and combined methods. Results are shown in $\#\mathcal{D}_{idi}$.}
    \vspace{-0.1in}
    \label{fig:combine-attack}
\end{figure}

\textbf{Scalability}. In our main experiments, we directly compared our \tool with the other two-phase baselines. Here, we further investigate the potential of our \tool as a fast global prober. In other words, we use \tool as a global seed generator to generate a seed set, and then combine it with other local methods (\ie, gradient-based method, genetic algorithm) to generate more instances around seeds. Specifically, we set the search number in the global phase as 40K (1K $\times$ 40 iters), the global seed set constraint as 1K, the iteration in the local phase as 1K, and the target model is DNN. The results are shown in Figure \ref{fig:combine-attack}, where ``ADF+\toolns'' indicates the test process consisting of a global phase of \tool and a local phase of ADF. We could observe that, by combining our \tool with other local methods, we could generate more discriminatory instances, \ie, \tool enhances the generation quantity around \textbf{1.65} times on average.
We conjecture the reason is that \tool could find sufficient discriminatory instances in the global phase, therefore facilitating the local methods that are limited by insufficient seeds, which usually occurs when testing small datasets (\eg, Credit) or relying on an imperfect explainer.

\textbf{Generalizability}. Besides tabular data, we here intend to evaluate the generalizability of our approach to image data. Unlike tabular data, the attributes of the image (\eg, Gender, Smile, and BigLips) are difficult to modify directly from the input domain. Therefore, we slightly adjust the modification of protected attributes in our \tool framework as follows: (1) we first approximate another linear hyperplane $h_{p}$ which separates the binary protected attribute (\eg, separating Gender to male and female); (2) Based on $h_{p}$, we then modify the protected attribute of potential discriminatory candidates as follows
\begin{equation} 
\mathbf{z}_{i}^{'} = \mathbf{z}_{i} -2 \cdot (\mathbf{w}_{p}^{T}\mathbf{z}+b_{p}) \cdot \mathbf{w}_{p}, where \, i \in \{0,+,-\},
\end{equation} 
where $\mathbf{w}_{p}$ and $b_{p}$ represent the unit normal vector and the unit intercept of the hyperplane $h_{p}$ respectively. Moreover, $(\mathbf{z}_{i}^{'}$,$\mathbf{z}_{i})$ forms a test case pair, which is different in the protected attribute.

\begin{figure}[ht]
\centering
\includegraphics[width=\linewidth]{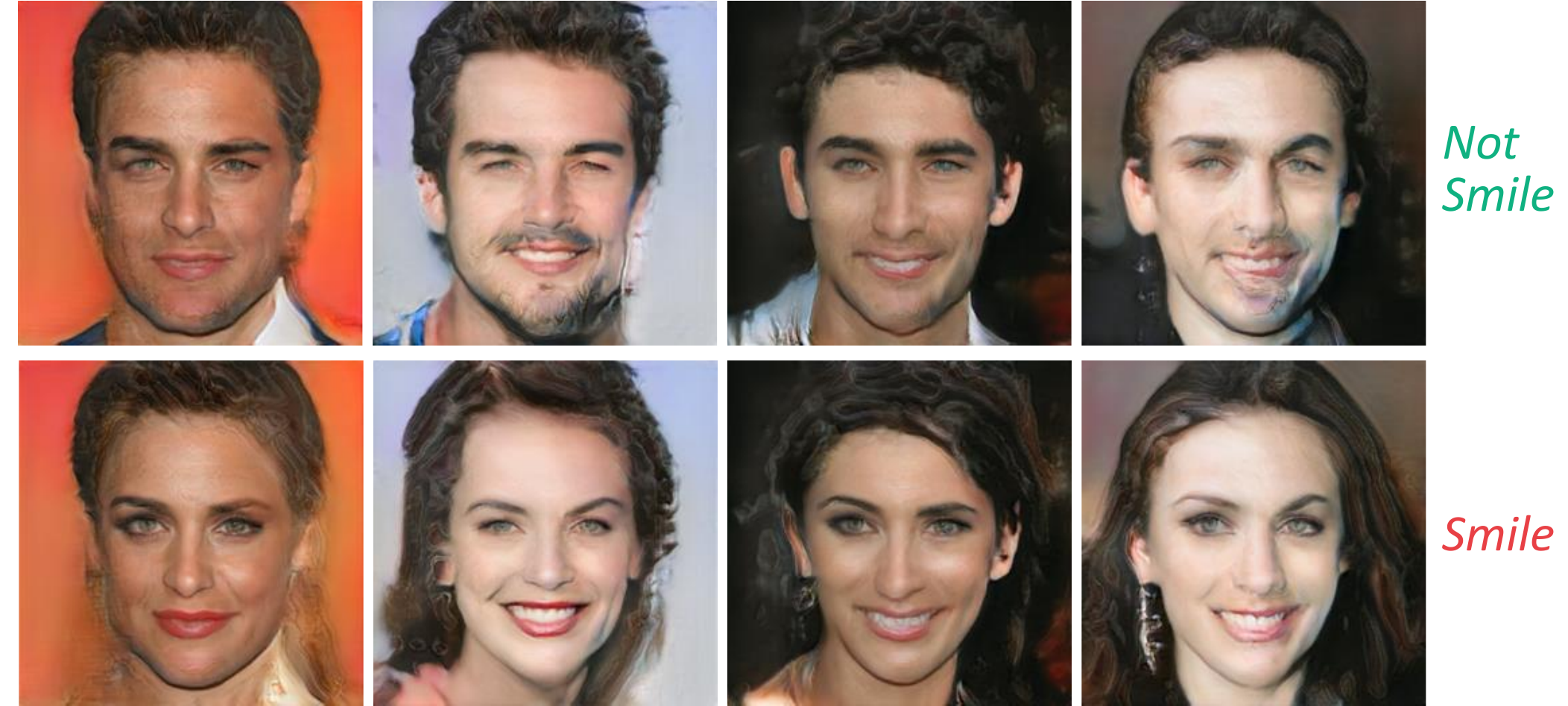}
\vspace{-0.3in}
\caption{Discriminatory images generated by \toolns. {Men are predicted with not smile (\emph{Top}), while women are predicted with smile (\emph{Bottom}).}}
\label{fig:cv_gender}
\vspace{-0.2in}
\end{figure}

In particular, we here test a ResNet-50 \cite{he2016deep}) model trained on the CelebA dataset \cite{liu2015deep} for smile detection, utilize a pre-trained Progressive GAN \cite{karras2017progressive} from GAN Zoo \cite{PGGAN}, and select the protected attribute as ``gender''. During one hour of testing, \tool obtains around 5K discriminatory samples that differ in gender from 50K test cases. As shown in Figure \ref{fig:cv_gender}, we can observe that \tool modifies the gender attribute obviously, and reveals the gender discrimination in the smile classifier which prefers giving smile predictions to the face of a female.

\section{Threats to Validity}

\quad\textbf{A well-trained GAN.} Our \tool relies on a well-trained GAN, such as CTGAN \cite{xu2019modeling} and Progressive GAN \cite{karras2017progressive}, to probe sufficient test cases in the latent space. For tabular data, we can easily obtain a well-trained GAN in just a few minutes; while for image data, due to the high dimensionality and unstructured nature, it is relatively difficult and requires more time. However, it is convenient for testers to access off-the-shelf GANs for testing in the image domain since the pre-trained weights are widely open-sourced \cite{PGGAN}.

\textbf{Protected attribute.} 
\revised{We follow the most common settings \cite{aggarwal2019black,zhang2020white,zhang2021efficient,zheng2022neuronfair,fan2022explanation,udeshi2018automated} and only consider the single protected attribute for each fairness testing in our main experiment.} However, testing for multiple protected attributes at a time will not hamper the performance of \toolns, but will certainly consume more time since all the possible combinations of the protected attributes perturbation need to be attempted. 
\revised{We here also study the combination of multiple protected attributes (gender\&race, gender\&age, race\&age on Adult; gender\&age on Credit), where our \tool still outperforms others (average results over datasets are shown in Table \ref{tab:combination-dnn}). More detailed results can be found on our website \cite{ourweb}.
}
\begin{table}[ht]
\caption{\revised{The combination of multiple protected attributes against DNN models. Average results over datasets are shown in $\#\mathcal{D}_{idi}$, $EGS$, and $ATN$ respectively.}}
\vspace{-0.1in}
  \label{tab:combination-dnn}
\resizebox{\columnwidth}{!}{%
\begin{tabular}{@{}r|cccccc@{}}

\toprule
\revised{Metric}   & \revised{AEQUITAS}     & \revised{ADF}     & \revised{EIDIG}   & \revised{NeuronFair} & \revised{ExpGA}  & \revised{LIMI}   \\ \midrule
    $\#\mathcal{D}_{idi}$$\textcolor{red}{\uparrow}$ & 1,752  & 10,158  & 8,972   & 9,897      & 23,422 & \pmb{120,164} \\
$EGS$$\textcolor{red}{\uparrow}$     & 0.49    & 2.82    & 2.49    & 2.75       & 6.51   & \pmb{33.38} \\
$ATN$$\textcolor{red}{\uparrow}$  & 50.85\% & 67.21\% & 65.41\% & 64.87\%    & 53.81\% & \pmb{82.39\%} \\ \bottomrule
\end{tabular}
}
\vspace{-0.1in}
\end{table}

\revised{
    \textbf{Generalization to textual data.} 
    In this paper, we follow the commonly-used settings \cite{udeshi2018automated,aggarwal2019black,zhang2020white,zhang2021efficient,zheng2022neuronfair} and primarily verify the effectiveness of our method on tabular data. Our method can be extended to other domains, such as the studies in Section \ref{sec:discussion} demonstrate the potential and generalizability to the image data. However, it is non-trivial to simply extend our current framework to textual data, since training GANs for text generation \cite{yu2017seqgan,zhou2020self} presents unique challenges owing to the sequential nature of textual data. The training challenge often results in mode collapse \cite{metz2016unrolled}, thus the generator would produce limited output and fail to capture the full range of possible outputs resulting in the weak ability to generate natural samples. However, we are still interested in exploring the possibilities of \tool in generalizing to Natural Language Processing (NLP) tasks in the future.
}

\vspace{-0.1in}
\section{Related Work}
\vspace{-0.05in}
\subsection{Fairness Testing} 
\vspace{-0.05in}
Deep learning models easily exhibit some undesirable behaviors on concerns such as robustness, privacy, and other trustworthiness issues \cite{Liu2023Xadv,Liu2019Perceptual,liu2021ANP,guo2023towards,Liu2020Spatiotemporal,Liu2020Biasbased,zhang2020interpreting,Wang2021DualAttention,tang2021robustart}. To reveal the discordance between existing and required fairness conditions of a given software system \cite{chen2022fairness}, a long line of work has been dedicated to performing model fairness testing such as test input generation \cite{xie2020fairness,black2020fliptest,perera2022search,udeshi2018automated,aggarwal2019black,zhang2020white,zhang2021efficient,zheng2022neuronfair,fan2022explanation,sharma2021mlcheck,diaz2018addressing} and test oracle identification \cite{hardt2016equality,barr2014oracle,rajan2022aequevox,sharma2019testing,chakraborty2020fairway,hort2021fairea,chakraborty2021bias,barocas2016big}. In this paper, we primarily focus on generating \emph{individual discriminatory instances} for machine learning model fairness testing, which can be roughly divided into white-box and black-box fairness testing based on the access to the target model.

For \textbf{black-box fairness testing}, testers have limited or even without any knowledge of the internal working of ML model (\eg, model architecture and gradients). Galhotra \etal \cite{galhotra2017fairness} first formally defined software fairness and discrimination, then proposed a fairness test method named THEMIS, which randomly generates test cases to measure the software discrimination. However, the random test input generation in THEMIS is inefficient. To accelerate the generation, Udeshi \etal \cite{udeshi2018automated} proposed AEQUITAS, a two-phase search-based generating approach. In global search phase, AEQUITAS also randomly samples the input space to discover the discriminatory inputs as test seeds. While in local phase, AEQUITAS designs three different strategies to explore the neighborhood of global seeds systematically, which directs the probability of the attributes to be perturbed. Subsequently, Agarwal \etal \cite{aggarwal2019black} presented SG, which combines symbolic execution and local explainability to detect individual discrimination. SG first utilizes the local explainer like LIME to construct a decision tree path to approximate the model decision process, and then leverages the symbolic execution to cover different tree paths to generate test cases. The ExpGA \cite{fan2022explanation} also uses the interpretable model to search for high-quality test seeds in the global phase, and employs the genetic algorithm to generate a large amount of discriminatory offspring rapidly. 

For \textbf{white-box fairness testing}, testers have complete knowledge of the target model and can fully access it. For example, Zhang \etal \cite{zhang2020white} proposed ADF, which firstly deals with the fairness testing problem of DNNs. ADF also contains two search phases, and requires the gradient information to search discriminatory instances near the decision boundary of DNN. Following the framework of ADF, Zhang \etal \cite{zhang2021efficient} proposed EIDIG, achieving better efficiency by integrating a momentum term in global phase and exploiting the prior information of gradient in local phase. Later, Zheng \etal \cite{zheng2022neuronfair} proposed NeuronFair, further improving the performance by only calculating the gradients of biased neurons that are identified by NeuronFair rather than the whole model. Due to the dependence on gradient information of model architecture, these methods cannot deal with the fairness testing of traditional ML models, which significantly limits their practical application. 

Though these methods have shown progress in fairness testing, there is no guarantee that the generated instances are legitimate or natural. In other words, existing techniques consider the generated instances effective as long as they can flip the predicted outcome after changing protected attribute, which may fail to obey the real-world constraints resulting in unnatural test samples (\eg, extreme values like 10-year-old children are authorized with the loan \revised{\cite{chen2022fairness}}). In contrast, this paper proposes the \tool framework to generate natural individual discriminatory instances \revised{by probing nearby the decision boundary} for black-box ML model fairness testing.

\revised{
We note that both DEEPJANUS \cite{riccio2020model} and \tool share a similar motivation of exploring nearby the decision boundary for natural samples, however, there are fundamental differences in our approaches and purposes. Our \tool is designed to derive a surrogate boundary in the latent space and probe potential discriminatory latent candidates, with the aim of generating discriminatory instances for individual fairness testing. In contrast, DEEPJANUS \cite{riccio2020model} employs an evolutionary algorithm within a model representation of the input domain to generate boundary cases for image classifiers, with the goal of characterizing the frontier of behaviors. 
}

\vspace{-0.25in}
\subsection{GAN-Based Software Testing}
\vspace{-0.05in}
Due to the strong generative capability, a variety of works have been proposed to explore the application of GAN in software testing. Some researchers directly utilize existing GANs to facilitate the testing process. Zhang \etal \cite{zhang2018deeproad} utilized GAN to synthesize driving scenes with various weather conditions in the metamorphic testing of autonomous driving systems. Similarly, Gao and Han \cite{gao2019automated} utilized DCGAN \cite{radford2015unsupervised} and CycleGAN \cite{zhu2017unpaired} for style transferring in coverage testing for deep learning systems. Rather than the image domain, Guo \etal \cite{guo2022automated} attempted three kinds of GANs to learn the execution path information of software in order to generate test data that can achieve full test coverage.
Other researchers propose variants of GAN to solve their problems specifically. Bao \etal \cite{bao2019actgan} proposed ACTGAN to capture the hidden structures of good configurations and generate potentially better configurations, which accelerates the configuration tuning process to a large extent. Porres \etal \cite{porres2021online} proposed an online GAN algorithm for automatic performance test generation, which can generate a high number of tests to reveal performance defects.

By contrast, this paper primarily focuses on black-box fairness testing, where we resort to the strong data-fitting ability of GANs to better generate natural individual discriminatory instances.

\vspace{-0.2in}
\section{Conclusion}
\vspace{-0.05in}
This paper proposes \tool framework to generate natural individual discriminatory instances for fairness testing. \tool first coarsely approximates the decision boundary of the tested model by deriving a surrogate linear boundary in the semantic latent space of GAN; \tool then manipulates the random latent vectors to the surrogate boundary with a one-step movement and further conduct vector calculation to probe two potential discriminatory candidates on either side of it. Therefore, we could generate individual discriminatory instances closer to the real decision boundary and thus \revised{acquire} better naturalness. Extensive experiments demonstrate that our \tool can generate a larger number of natural discriminatory instances with a higher speed than 6 SOTA methods in 7 benchmarks. Moreover, the model fairness can be improved by retraining with the natural discriminatory instances generated by \toolns.

\textbf{Acknowledgement.} This work was supported by the National Key R\&D Program of China (2022ZD0116310), the National Natural Science Foundation of China (62022009 and 62206009), and the State Key Laboratory of Software Development Environment.

\bibliographystyle{ACM-Reference-Format}
\bibliography{main}

\end{document}